\documentstyle[12pt,axodraw]{article}
\begin{document}
\baselineskip 0.6cm

\newcommand{\gsim}{\, \mathop{}_{\textstyle \sim}^{\textstyle >} \,}
\newcommand{\lsim}{\, \mathop{}_{\textstyle \sim}^{\textstyle <} \,}
\newcommand{\vev}[1]{ \langle {#1} \rangle }

\begin{titlepage}

\begin{flushright}
UCB-PTH-02/20 \\
LBNL-50199 \\
\end{flushright}

\vskip 0.5cm

\begin{center}
{\Large \bf A Complete Theory of Grand Unification \\ 
 in Five Dimensions}

\vskip 0.6cm

{\large
Lawrence J.~Hall and Yasunori Nomura
}

\vskip 0.3cm

{\it Department of Physics, University of California,
                Berkeley, CA 94720, USA}\\
{\it Theoretical Physics Group, Lawrence Berkeley National Laboratory,
                Berkeley, CA 94720, USA}

\vskip 0.6cm

\abstract{A fully realistic unified theory is constructed, with $SU(5)$ 
gauge symmetry and supersymmetry both broken by boundary conditions in 
a fifth dimension. Despite the resulting explicit breaking of $SU(5)$ 
locally at a boundary of the dimension, when the size of the extra 
dimension is taken to be large precise predictions emerge for gauge 
coupling unification, $\alpha_s(M_Z) = 0.118 \pm 0.003$, and for Yukawa 
coupling unification, $m_b(M_Z) = 3.3 \pm 0.2~{\rm GeV}$. The 5D theory 
is then valid over a large energy interval from the compactification 
scale, $M_c \simeq 1 \times 10^{15}~{\rm GeV}$, to the scale of 
strong coupling, $M_s \simeq 1 \times 10^{17}~{\rm GeV}$. A complete 
understanding of the Higgs sector of the minimal supersymmetric 
standard model is given; with explanations for why the Higgs triplets 
are heavy, why the Higgs doublets are protected from a large tree-level 
mass, and why the $\mu$ and $B$ parameters are naturally generated to 
be of order the supersymmetry breaking scale. All sources of proton 
decay from operators of dimension four and five are forbidden, while 
a new origin for baryon number violating dimension six operators is 
found to be important. The exchange of the superheavy gauge boson, 
with a brane-localized kinetic energy interaction, leads to 
$\tau_p \approx 10^{34}~{\rm years}$, with several branching ratios 
determined in terms of a single mixing parameter. The theory is only 
realistic for an essentially unique choice of matter location in the 
fifth dimension: the ten-plets of the first two generations must lie 
in the bulk, with all other matter located on the $SU(5)$ preserving 
boundary. Several aspects of flavor follow from this geometry: only 
the third generation possesses an $SU(5)$ mass relation, and the 
lighter two generations have only small mixings with the heaviest 
generation except for neutrinos. The entire superpartner spectrum is 
predicted in terms of only two free parameters. The squark and slepton 
masses have sizes determined by their location in the fifth dimension, 
allowing a significant experimental test of the detailed structure of 
the extra dimension. Lepton flavor violation is found to be generically 
large in higher dimensional unified theories with non-trivial matter 
geometries, providing soft supersymmetry breaking operators are local 
up to the compactification scale. In our theory this forces a common 
location for all three neutrinos, predicting large neutrino mixing 
angles. Rates for $\mu \rightarrow e \gamma$, $\mu \rightarrow e e e$, 
$\mu \rightarrow e$ conversion and $\tau \rightarrow \mu \gamma$ are 
larger in our theory than in conventional 4D supersymmetric grand 
unified theories, and, once superpartner masses are measured, these 
rates are completely determined in terms of two leptonic mixing angles. 
Proposed experiments probing $\mu \rightarrow e$ transitions will 
probe the entire interesting parameter space of our theory.}

\end{center}
\end{titlepage}

\section{Introduction}
\label{sec:intro}

Weak scale supersymmetry not only provides a framework for electroweak 
symmetry breaking, but also leads to a highly successful prediction 
for the unification of gauge couplings.  If this picture of a 
supersymmetric desert is correct, the scale of gauge coupling 
unification heralds the threshold for some new unified physics; 
for example, conventional supersymmetric grand unified theories or 
string theory. We have recently introduced a new alternative for 
the unified physics, which we call Kaluza-Klein (KK) grand 
unification \cite{Hall:2001pg, Hall:2001xb}.  In this framework the 
grand unified symmetry is realized in higher dimensions, but is 
explicitly broken locally by defect branes, and consequently does not 
appear as a symmetry of the low energy effective 4D theory. In general 
the gauge symmetry breaking associated with the local defects destroys 
gauge coupling unification; however, if the volume of the bulk is 
large, this symmetry breaking is diluted, restoring gauge coupling 
unification \cite{Hall:2001pg}.  In all known 4D grand unified 
theories, the accuracy of the prediction for the QCD coupling from 
gauge coupling unification is limited because the unified threshold 
corrections cannot be computed. In KK grand unification, if the volume 
of the bulk is increased so that the theory becomes strongly coupled 
at the cutoff scale, the leading unified scale corrections can be 
computed, leading to a new level of precision for gauge coupling 
unification \cite{Hall:2001xb}.

Several features of KK grand unified theories make them extremely 
attractive candidates for physics in the region of the unification 
scale. They incorporate the advances of conventional grand unified 
theories, such as charge quantization and the quantum numbers of 
the quarks and leptons, while overcoming their problematic features. 
In particular the orbifold boundary conditions automatically require 
that multiplets in the bulk are split in mass. This is particularly 
important for the gauge and Higgs multiplets, and provides an elegant 
origin for gauge symmetry breaking and for a large mass splitting 
between the Higgs triplets and doublets \cite{Kawamura:2001ev}. 
While the simplest 4D supersymmetric $SU(5)$ theory 
\cite{Dimopoulos:1981zb} is excluded, by too large a proton decay 
rate mediated by colored Higgsino exchange \cite{Sakai:1981pk}, in 
KK grand unified theories the form for the Higgsino mass matrix is 
determined by the KK mode expansion and automatically forbids proton 
decay from Higgsino exchange \cite{Hall:2001pg}.  Finally, KK grand 
unification does not lead to fermion mass relations for all matter 
--- no relations are expected for bulk matter \cite{Hall:2001pg, 
Hebecker:2001wq}. Hence there is a successful correlation: only the 
heavier fermions are expected to exhibit unified mass relations 
\cite{Hall:2001xb, Hall:2001zb, Nomura:2001tn}.

In Ref.~\cite{Hall:2001xb} we have constructed a minimal theory of 
KK grand unification, which has $SU(5)$ gauge symmetry in 5D and 
provides a uniquely successful, high-precision prediction for the QCD 
coupling, $\alpha_s(M_Z) = 0.118 \pm 0.004 \pm 0.003$, where the first 
uncertainty arises from the supersymmetric threshold and the second from 
the scale of strong coupling. It is only in this case that the unified 
scale corrections, from the KK towers of gauge and minimal Higgs 
sectors, give agreement with experiment $\alpha_s^{\rm exp}(M_Z) = 
0.117 \pm 0.002$ \cite{Groom:2000in}. In this theory $R$ parity arises 
as a subgroup of a continuous $U(1)_R$ symmetry that is related to the 
$SU(2)_R$ symmetry of the bulk supersymmetry. This $U(1)_R$ symmetry 
forbids a mass operator for the Higgs fields, completing the solution 
to the doublet-triplet splitting problem, and forbids all proton decay 
from operators of dimension four and five.  Furthermore, this theory is 
sufficiently tight that certain aspects of flavor must be related to 
the geometrical location of matter in the extra dimension. In particular
the top quark resides on a brane while the up quark is in the bulk. 

In this paper we pursue this 5D $SU(5)$ theory further,
addressing two questions:
\begin{itemize}
\item Can the theory be made completely realistic?
\item Can the theory be experimentally tested?
\end{itemize}
These two questions are closely related. Supersymmetry breaking is the 
major remaining additional ingredient needed for the theory to be fully
realistic, and it is via the precise form of the soft supersymmetry 
breaking interactions that further tests of the theory are possible. 
Clearly there may be several ways to successfully incorporate 
supersymmetry breaking, and hence several versions of the fully 
realistic theory to test. In this paper we break supersymmetry by 
boundary conditions, using the same extra dimension that breaks the 
gauge symmetry \cite{Barbieri:2001yz}. This is highly economical, 
involving the vacuum expectation value of a field in the 5D gravity 
multiplet \cite{Marti:2001iw}, and is highly predictive, since the most 
general such boundary condition involves just a single free parameter 
$\alpha$. Such supersymmetry breaking further constrains the location 
of matter because squarks and sleptons in the bulk acquire a tree-level 
mass $\tilde{m} = \alpha/R$, while those on the brane are massless at 
tree level. Combining previous constraints on matter location with 
considerations of superpartner induced flavor changing interactions, 
the location of every quark and lepton is determined, up to a two-fold 
ambiguity. In particular, the three $SU(5)$ five-plets, $F_i$, must 
either all be on the $SU(5)$ preserving brane, as shown in 
Fig.~\ref{fig:theory}, or they must all be in the bulk. We will
concentrate on the first case, since only then does a unified 
prediction for $m_b / m_\tau$ result. The second case is not 
uninteresting, since it leads to a geometric suppression of $m_b/m_t$, 
although we will find that some of this ratio must originate in 
$\tan\beta$. 

Although we provide numerical predictions for this particular origin 
of supersymmetry breaking, many of the physical effects and signals 
we consider in this paper are much more general.  These include 
contributions to $m_b/m_\tau$ from KK towers, an essential uniqueness 
of the matter location and its consequence of large mixing angles for 
neutrinos, superpartner masses reflecting the geometry of their 
locations, large lepton flavor violating signals, and a new origin and 
predictions for gauge boson mediated proton decay.  In particular, 
we propose lepton flavor violation as a powerful and generic signal for 
KK grand unified theories with high mediation scales of supersymmetry 
breaking, and identify the general structure of soft supersymmetry 
breaking operators at the compactification scale by studying the 
flavor symmetry of the 5D gauge interactions.

The entire spectrum for the superpartners and the Higgs sector is
predicted in our theory in terms of the supersymmetry breaking mass
scale $\tilde{m}$ and the ratio of electroweak vacuum expectation 
values, $\tan\beta$. The predictions are characteristic of the 
underlying locations of each matter field in the extra dimension, and 
therefore provide a significant probe of the short distance structure 
of the theory. The combination of matter location and boundary 
condition supersymmetry breaking leads to flavor violation in the 
superpartner interactions. Large lepton flavor violating signals are 
predicted in terms of the parameters $\tilde{m}$, $\tan \beta$ and 
two flavor mixing angles $\theta^e_{12}$ and  $\theta^e_{23}$. Future
experiments probing $\mu \rightarrow e$ and $\tau \rightarrow \mu$
transitions could probe essentially all of the parameter space of the
theory where electroweak symmetry breaking occurs naturally. Finally
we give predictions from coupling constant unification for both
$\alpha_s(M_Z)$ and $m_b(M_Z)$ which include corrections from KK 
towers at the unified scale and from superpartners at the weak scale.

Neutrino masses occur in our theory via the see-saw mechanism 
\cite{Seesaw}.  The masses of the right-handed neutrinos are governed 
by the breaking of a $U(1)_X$ gauge symmetry near the compactification 
scale. The neutrino flavor mixing angles are expected to be large 
because the neutrinos reside in the five-plets $F_i$, which all have 
a common location. Our theory contains a brane coupling between the 
two Higgs doublets and a gauge singlet field $X$ of the form 
$X H \bar{H}$. The supersymmetric dynamics which breaks $U(1)_X$ gauge 
symmetry determines $X$ to have vanishing vacuum expectation values. 
However, once supersymmetry breaking is included, a readjustment of 
the vacuum occurs so that $\vev{X} \approx \tilde{m}$ and $\vev{F_X} 
\approx \tilde{m}^2$, providing a natural origin for $\mu$ and $\mu B$ 
parameters \cite{Hall:2002up}.

In sub-section \ref{subsec:theory1} we review the basic features of our 
$SU(5)$ theory in 5D, paying particular attention to gauge coupling 
unification and $U(1)_R$ symmetry.  In sub-section \ref{subsec:theory2} 
we introduce boundary condition supersymmetry breaking, and give 
the form for the soft operators and the predictions for the soft 
supersymmetry breaking parameters at the weak scale.  In section 
\ref{sec:unif} we discuss several consequences of our theory: quark 
and lepton masses, supersymmetric threshold corrections to gauge 
coupling unification, Yukawa coupling unification, proton decay, and 
$R$ axions and axinos.  While we work in the specific context of 
boundary condition supersymmetry breaking, some of our analyses, 
for example for proton decay and unified scale correction to $b/\tau$ 
unification, are completely independent of how supersymmetry is broken.
In section \ref{sec:flavor} we study the general structure of flavor 
symmetries in KK grand unified theories and argue that large lepton 
flavor violation is a generic signature of these theories, providing 
that soft supersymmetry breaking operators are generated at or above 
the compactification scale.  We then study the supersymmetric flavor 
violation induced in our theory, paying particular attention to the 
predictions in the lepton sector.  In section \ref{sec:develop} 
we discuss neutrino masses and the generation of the $\mu$ term, which 
are linked by the breaking of $U(1)_X$ gauge symmetry. Finally, in 
section \ref{sec:alt} we discuss the variant of our theory where the 
$F_i$ are located in the bulk.  Conclusions are drawn in 
section \ref{sec:concl}.

\section{The Theory}
\label{sec:theory}

In this section we introduce our theory.  In sub-section 
\ref{subsec:theory1}, we overview the 5D $SU(5)$ KK grand unified 
theory of Ref.~\cite{Hall:2001xb}, discussing the symmetry structure 
and field content. We show that gauge coupling unification occurs,
despite a point symmetry defect, and is in precise agreement with data. 
We explain solutions to  the three outstanding problems of 4D 
supersymmetric grand unification: doublet-triplet splitting, proton 
decay, and fermion mass relations. We find that an $R$ symmetry, 
originating from the 5D supersymmetry, is crucial for a successful 
phenomenology.

In sub-section \ref{subsec:theory2}, we introduce supersymmetry 
breaking by boundary conditions following Ref.~\cite{Barbieri:2001yz}, 
leading to the usual soft supersymmetry breaking operators with 
coefficients determined by a single free parameter.  Predictions for 
the supersymmetric particle spectrum are given, together with 
a brief discussion of collider phenomenology.

\subsection{Minimal Kaluza-Klein grand unification}
\label{subsec:theory1}

\subsubsection{Boundary conditions and restricted unified gauge symmetry}

We consider a 5D $SU(5)$ supersymmetric gauge theory 
compactified on an $S^1/Z_2$ orbifold.  The 5D gauge multiplet 
${\cal V} = \{ A_M, \lambda, \lambda', \sigma \}$ consists of a 5D 
vector field, $A_M$, two gauginos, $\lambda$ and $\lambda'$, and a real 
scalar, $\sigma$.  Compactification on $S^1/Z_2$ is obtained by 
identifying the fifth coordinate $y$ under the two operations, 
${\cal Z}: y \rightarrow -y$ and ${\cal T}: y \rightarrow y + 2\pi R$.  
The resulting space is a line interval $y \in [0,\pi R]$, with 
boundaries at $y=0$ and $\pi R$. Boundary conditions are chosen 
so that the orbifold reflection ${\cal Z}$ reduces 5D $N=1$ 
supersymmetry to 4D $N=1$ supersymmetry and preserves $SU(5)$, while 
the translation ${\cal T}$ breaks $SU(5)$ by the action of 
$P = \mbox{diag}(+,+,+,-,-)$ on a 5-plet.\footnote{
This is equivalent to the boundary conditions of 
Ref.~\cite{Kawamura:2001ev} described in terms of ${\cal Z}$ and 
${\cal Z}' = {\cal Z}{\cal T}$ as $S^1/(Z_2 \times Z_2')$.}
In particular, the boundary conditions of the gauge multiplet are 
given by 
\begin{equation}
  \pmatrix{V^{(p)} \cr \Sigma^{(p)}}(x^\mu,y) 
  = \pmatrix{V^{(p)} \cr -\Sigma^{(p)}}(x^\mu,-y) 
  = p \pmatrix{V^{(p)} \cr \Sigma^{(p)}}(x^\mu,y+2\pi R),
\label{eq:bc-g}
\end{equation}
where we have used the 4D $N=1$ superfield language, 
${\cal V} = \{ V, \Sigma \}$: $V(A_\mu, \lambda)$ and 
$\Sigma((\sigma+iA_5)/\sqrt{2}, \lambda')$ are 4D vector and chiral 
superfields in the adjoint representation.  The standard model gauge 
multiplets $(V_{321},\Sigma_{321}) \equiv (V^{(+)},\Sigma^{(+)})$
have positive eigenvalues for the $P$ matrix, $p = 1$, while the broken 
$SU(5)$ gauge multiplets $(V_X,\Sigma_X) \equiv (V^{(-)},\Sigma^{(-)})$ 
have negative eigenvalues, $p=-1$.  After the KK decomposition, only 
the minimal supersymmetric standard model (MSSM) gauge multiplets, 
$V_{321}$, have massless modes, and all the other modes have masses of 
the order of the compactification scale $M_c \equiv 1/R$, as summarized 
in Table~\ref{table:Z-T}.  We will see later that $M_c$ must be very 
large, of order $10^{15}~{\rm GeV}$, but it differs from the conventional 
unification mass scale $M_u \simeq 2 \times 10^{16}~{\rm GeV}$.
\begin{table}
\begin{center}
\begin{tabular}{|c|c|c|c|c|}
\hline
 $({\cal Z},{\cal T})$  &  gauge and Higgs fields  & 
    bulk matter fields & KK modes & 4D masses \\ \hline
 $(+,+)$  & $V_{321}$,      $H_D$,   $\bar{H}_D$   & 
    $T_{U,E}$, $T'_Q$,     $F_D$, $F'_L$      & 
    $\cos[ny/R]$       & $n/R$       \\ 
 $(+,-)$  & $V_{X}$,        $H_T$,   $\bar{H}_T$   & 
    $T_Q$, $T'_{U,E}$,     $F_L$, $F'_D$      & 
    $\cos[(n+1/2)y/R]$ & $(n+1/2)/R$ \\ 
 $(-,+)$  & $\Sigma_{321}$, $H^c_D$, $\bar{H}^c_D$ & 
    $T^c_{U,E}$, $T'^c_Q$, $F^c_D$, $F'^c_L$  & 
    $\sin[(n+1)y/R]$   & $(n+1)/R$   \\ 
 $(-,-)$  & $\Sigma_{X}$,   $H^c_T$, $\bar{H}^c_T$ & 
    $T^c_Q$, $T'^c_{U,E}$, $F^c_L$, $F'^c_D$  & 
    $\sin[(n+1/2)y/R]$ & $(n+1/2)/R$ \\ 
\hline
\end{tabular}
\end{center}
\caption{The transformation properties for the bulk fields 
 under the orbifold reflection and translation.  Here, we have used 
 the 4D $N=1$ superfield language.  The fields written in the 
 $({\cal Z},{\cal T})$ column, $\varphi$, obey the boundary condition 
 $\varphi(y) = {\cal Z}\varphi(-y) = {\cal T}\varphi(y+2\pi R)$.
 The modes and masses for the corresponding KK towers are also given 
 ($n=0,1,\cdots$).}
\label{table:Z-T}
\end{table}

\begin{figure}
\begin{center}
    \input{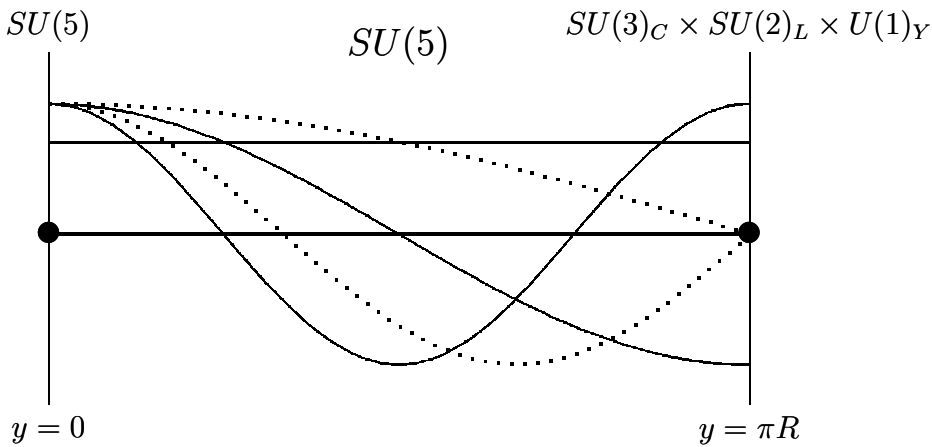}
\caption{In the fifth dimension, space is a line segment bounded by 
 branes at $y=0$ and at $y=\pi R$.  Here, solid and dotted lines 
 represent the profiles of gauge transformation parameters $\xi_{321}$ 
 and $\xi_X$, respectively.  Because $\xi_X(y = \pi R) = 0$, 
 explicit point defect symmetry breaking occurs at the $y = \pi R$ 
 brane, which only respects $SU(3)_C \times SU(2)_L \times U(1)_Y$ 
 gauge symmetry.}
\label{fig:orbifold}
\end{center}
\end{figure}

What is the gauge symmetry of this theory?  While the low-energy 4D 
theory has only the standard model gauge symmetry, the original 5D 
theory has a larger gauge symmetry.  We find that this gauge symmetry 
is $SU(5)$ but with the gauge transformation parameters obeying the 
same boundary conditions as the corresponding 4D gauge fields:
\begin{equation}
  \xi^{(p)}(x^\mu,y) 
    = \xi^{(p)}(x^\mu,-y) 
    = p\, \xi^{(p)}(x^\mu,y+2\pi R),
\label{eq:bc-xi}
\end{equation}
which we refer to as restricted gauge symmetry \cite{Hall:2001pg}. 
The KK expansions for the standard model gauge parameters, $\xi^{(+)} 
= \xi_{321}$, and $SU(5)/(SU(3)_C \times SU(2)_L \times U(1)_Y)$ 
ones, $\xi^{(-)} = \xi_X$, are
\begin{eqnarray}
  \xi_{321}(x^\mu,y) &=& 
    \sum_{n=0}^\infty \xi_{321}^n(x^\mu) \cos{ny \over R},
\\
  \xi_X(x^\mu,y)     &=& 
    \sum_{n=0}^\infty \xi_X^n(x^\mu)\cos{(n+1/2)y \over R}.
\label{eq:xi}
\end{eqnarray}
Since $\xi_X$ always vanish at $y=\pi R$, the gauge symmetry is 
reduced to $SU(3)_C \times SU(2)_L \times U(1)_Y$ on this point, 
while the full $SU(5)$ symmetry is operative in all the other 
points in the extra dimension, as depicted in Fig.~\ref{fig:orbifold}.  

This structure allows the introduction of three types of fields: 4D 
$N=1$ superfields localized on the $y=0$ brane in representations of 
$SU(5)$, 4D $N=1$ superfields on the $y=\pi R$ brane in representations 
of the standard model gauge group, and bulk fields forming 5D $N=1$ 
supermultiplets and representations of $SU(5)$. In particular, we can 
introduce a hypermultiplet ${\cal H}_\Phi = \{ \phi, \phi^c, \psi, 
\psi^c \}$ in the bulk, which consists of two complex scalars, $\phi$ 
and $\phi^c$, and two Weyl fermions, $\psi$ and $\psi^c$.  The boundary 
conditions for hypermultiplets are given by 
\begin{equation}
  \pmatrix{\Phi^{(p)} \cr \Phi^{c(p)}}(x^\mu,y) 
  = \pmatrix{\Phi^{(p)} \cr -\Phi^{c(p)}}(x^\mu,-y) 
  = p\, \eta_{\Phi} 
    \pmatrix{\Phi^{(p)} \cr \Phi^{c(p)}}(x^\mu,y+2\pi R), 
\label{eq:bc-h}
\end{equation}
where we have used the 4D $N=1$ superfield language, 
${\cal H}_\Phi = \{ \Phi, \Phi^c \}$: $\Phi(\phi, \psi)$ and 
$\Phi^c(\phi^c, \psi^c)$ are 4D chiral superfields. 
An overall factor $\eta_\Phi$ appears in the orbifold translation, 
which can be chosen to be $\eta_\Phi = \pm 1$ for each hypermultiplet.

The two Higgs doublets of the MSSM are introduced in the bulk as 
two hypermultiplets ${\cal H}_H = \{ H, H^c \}$ and 
${\cal H}_{\bar{H}} = \{ \bar{H}, \bar{H}^c \}$, which transform as 
${\bf 5}$ and $\bar{\bf 5}$ under $SU(5)$.  The boundary conditions 
are given by Eq.~(\ref{eq:bc-h}) with $\eta_H = \eta_{\bar{H}} = -1$, 
so that we have massless Higgs doublets.  (In the present notation, 
$H^{(+)}$ and $H^{(-)}$ ($\bar{H}^{(+)}$ and $\bar{H}^{(-)}$) represent 
triplet and doublet components, $H_T$ and $H_D$ ($\bar{H}_T$ and 
$\bar{H}_D$), of $H$ ($\bar{H}$), respectively.)  The resulting KK 
towers are summarized in Table~\ref{table:Z-T}.  These Higgs KK 
towers do not have zero modes for the color triplet states 
\cite{Kawamura:2001ev}.  Moreover, since the mass term of the KK 
excitations takes the form $H H^c + \bar{H} \bar{H}^c$ rather than 
$H \bar{H}$, the exchange of triplet states does not lead to 
proton decay from dimension five operators \cite{Hall:2001pg}. 
At low energies, only the two Higgs doublets of the MSSM, 
$H_D$ and $\bar{H}_D$, remain from the Higgs KK towers.

The quarks and leptons are introduced either in the bulk or on the 
$SU(5)$ brane at $y=0$, to preserve the $SU(5)$ understanding 
of matter quantum numbers \cite{Hall:2001pg}.  If quarks 
and leptons are on the brane, they fill out 4D chiral multiplets 
which are ${\bf 10}$ or $\bar{\bf 5}$ representations of $SU(5)$: 
$T$ and $F$.  On the other hand, if quarks and leptons are in the bulk, 
they arise from hypermultiplets: $\{ T,T^c \} + \{ T',T'^c \}$ 
with $\eta_T=-\eta_{T'}=1$ and $\{ F,F^c \} + \{ F',F'^c \}$ with 
$\eta_F=-\eta_{F'}=1$.  We then find from Eq.~(\ref{eq:bc-h}) that 
a generation $q,u,d,l,e$ arises from the zero modes of bulk 
fields $T(u,e), T'(q), F(d)$ and $F'(l)$.  (The tower structure 
for these fields is given in Table~\ref{table:Z-T}. Note that 
$T^{(+)} = T_{U,E}$, $T^{(-)} = T_{Q}$, $F^{(+)} = F_{D}$, 
$F^{(-)} = F_{L}$, and similarly for $T'$ and $F'$, where $T_{Q,U,E}$ 
($F_{D,L}$) are the components of $T$ ($F$) decomposed into irreducible 
representations of the standard model gauge group.)  We can choose 
where we put quarks and leptons for each $SU(5)$ representation, 
${\bf 10}$ and $\bar{\bf 5}$, in each generation.  Thus, at this stage, 
we have $(2^2)^3 = 64$ different choices for the configuration 
of matter.  We will identify the most attractive matter configuration 
later in this section.

\subsubsection{Gauge coupling unification from strong coupling}

So far, we have demonstrated how the orbifold compactification of
the 5D $SU(5)$ theory leads to the gauge group and matter content of  
the MSSM in the 4D effective theory below $M_c$. However, since 
$SU(5)$ is explicitly broken by boundary conditions, it is not obvious 
that this theory preserves successful gauge coupling unification.
In fact, we find that gauge coupling unification is generically 
destroyed due to the presence of local $SU(5)$ breaking on the 
$y = \pi R$ brane.  To see this, let us consider the effective field 
theory above $M_c$.  Since the higher dimensional gauge theory is 
non-renormalizable, this effective theory must be cut off at some 
scale $M_s$, where the theory is embedded into a more fundamental theory 
such as string theory.  At the scale $M_s$, the most general effective 
action for the gauge kinetic terms is
\begin{equation}
  S = \int d^4x \; dy \; 
    \biggl[ \frac{1}{g_5^2} F^2 + 
    \delta(y) \frac{1}{\tilde{g}^2} F^2 +
    \delta(y - \pi R) \frac{1}{\tilde{g}_a^2} F_a^2 \biggr],
\label{eq:gaugekinops}
\end{equation}
where the first and second terms are $SU(5)$-invariant bulk and brane 
gauge kinetic energies, while the third term represents non-unified 
kinetic operators located on the $y=\pi R$ brane ($a = 1,2,3$ 
represents the standard model gauge groups).  This form is ensured by 
the restricted gauge symmetry, regardless of the unknown ultraviolet 
physics above $M_s$.  The standard model gauge couplings in the 
equivalent KK theory, $g_a$, are obtained by integrating over the 
extra dimension:
\begin{equation}
  \frac{1}{g_a^2} = \frac{\pi R}{g_5^2} + \frac{1}{\tilde{g}_a^2},
\label{eq:4d-gi}
\end{equation}
where the contribution from $\tilde{g}$ has been absorbed into a shift 
of the $\tilde{g}_a$.  This shows that $g_a$ depend on the coefficients 
of the localized kinetic operators, $\tilde{g}_a$, and are not universal 
at the scale $M_s$. However, this difficulty is overcome by requiring 
the extra dimension to have a large volume \cite{Hall:2001pg}.  Writing 
$g_5^2 = \gamma/M_s$ and $\tilde{g}_a^2 = \gamma_a$, we find that 
the non-universal term is suppressed compared with the universal term 
by a volume factor $(\gamma/\gamma_a)(1/ \pi M_s R)$.  Therefore, 
by making the extra dimension large ($\pi R M_s$ large), gauge coupling 
unification is recovered.

How large should we take the extra dimension?  It depends on the 
unknown coefficients $\gamma$ and $\gamma_a$.  In the extreme case 
of $\gamma_a \ll \gamma$, we even cannot recover gauge coupling 
unification by making the extra dimension large. 
In Ref.~\cite{Hall:2001xb}, we have introduced a framework which 
removes these concerns and makes KK grand unification more reliable 
and predictive.  The crucial new ingredient is the assumption that 
the gauge interaction is strongly coupled at the cutoff scale 
$M_s$ \cite{Nomura:2001tn}. While the theory is weakly coupled 
at $M_c$, it becomes strongly coupled at higher energies since 
a higher dimensional gauge coupling has negative mass dimensions. 
In the 4D picture this follows because the loop expansion parameter 
of the theory (the strength of the gauge interaction) is given by the 
usual loop factor, $C(g^2/16\pi^2)$, times the number of KK states 
available at the energy $E$, $N_{\rm KK}(E) \simeq (E/M_c)$, where 
$C$ represents a group theoretical factor ($C \simeq 5$). 
We require this loop expansion parameter to be $1$ at $M_s$: 
$C(g^2/16 \pi^2)(M_s/M_c) \simeq 1$. Although $g$ itself also depends 
on the energy $E$, its evolution is slow up to energies very close to 
$M_s$ so that we may take $g$ to be the 4D gauge coupling at $M_c$, 
$g \simeq 0.7$, to estimate $M_s/M_c$, which gives 
$M_s/M_c \simeq 300/C$.  This strong coupling requirement has the 
following virtues.  First, it allows us to estimate $\gamma$ and 
$\gamma_a$ by requiring that all the loop diagrams contribute equally 
at the scale $M_s$.  By carefully evaluating expansion parameters, 
we find $\gamma \simeq 16\pi^3/C$ and $\gamma_a \simeq 16\pi^2/C_a$, 
excluding the unwanted situation $\gamma_a \ll \gamma$.  This argument 
is quite similar to the case of the usual chiral theory of mesons: 
all the operator coefficients at the QCD scale $\Lambda$ are estimated 
to be products of appropriate powers of $4\pi$ and $\Lambda$ by requiring 
that all the loop diagrams contribute equally \cite{Weinberg:1978kz}. 
Substituting these estimates into Eq.~(\ref{eq:4d-gi}), we obtain
\begin{equation}
  \frac{1}{g_a^2} \simeq \frac{C M_s R}{16\pi^2} + \frac{C_a}{16\pi^2},
\end{equation}
at the scale $M_s$, where $C \simeq C_a \simeq 5$.  To obtain 
$g_a \simeq 0.7$ requires $M_s R \simeq 60$, so that the non-unified 
contribution from unknown ultraviolet physics is suppressed to be 
a negligible level (less than a $1\%$ correction to $g_a$).  We adopt 
this strong coupling scenario in the rest of the paper.

\subsubsection{Consequences of the extra dimension being large}

The presence of a moderately large extra dimension, $M_s R = O(100)$, 
has several important consequences.  First of all, the running of the 
gauge couplings between $M_s$ and $M_c$ gives a non-negligible 
contribution to the prediction of the QCD coupling.  In the energy 
interval between $M_s$ and $M_c$, the gauge couplings receive both 
power-law and logarithmic contributions.  However, the leading power-law 
piece comes from the renormalization of the bulk gauge coupling and 
thus must be universal due to the restricted gauge symmetry.  On the 
other hand, the logarithmic contributions come from the runnings of 
4D gauge kinetic terms localized on the branes and can be different 
for $SU(3)_C$, $SU(2)_L$ and $U(1)_Y$.  This means that the differences 
of the three gauge couplings evolve logarithmically above the 
compactification scale, although the gauge couplings themselves  
receive power law corrections \cite{Hall:2001pg, Nomura:2001mf}.
Since the beta-function coefficients for the relative 
runnings above $M_c$ are different from the MSSM beta-function 
coefficients, the prediction of minimal KK grand unification 
for the QCD coupling, $\alpha_s^{\rm KK}$, is different from the 
prediction of the single scale unification, $\alpha_s^{\rm SGUT,0}$.
The difference $\delta\alpha_s \equiv \alpha_s^{\rm KK} - 
\alpha_s^{\rm SGUT,0}$ is given by
\begin{equation}
  \delta\alpha_s \simeq -\frac{3}{7\pi} \alpha_s^2 \ln\frac{M_s}{M'_c},
\label{eq:as-formula}
\end{equation}
where $M'_c \equiv M_c/\pi$ \cite{Hall:2001xb}. An important point is 
that $\delta\alpha_s$ is dominated by the calculable contribution 
coming from the energy interval between $M_s$ and $M'_c$, since it 
gives a non-universal correction to $1/g_a$ by an amount of order
$\simeq (C_a/16\pi^2)\ln(M_s/M'_c)$ which is larger than that from 
unknown ultraviolet physics, $\simeq (C_a/16\pi^2)$, by a factor 
of $\ln(M_s/M'_c)$.  The other uncertainties are also under control: 
the dependence of $\delta\alpha_s$ on $M_s/M'_c$ is weak, and the 
effect from the strong coupling physics around $M_s$ is 
small \cite{Nomura:2001tn}.  Therefore, we obtain 
$\delta\alpha_s \approx -0.01$ from $M_s/M'_c \approx 100$, 
eliminating the discrepancy between the usual supersymmetric 
prediction and data. The compactification scale $M'_c$ is given by
\begin{equation}
  M'_c = M_u \left( \frac{M'_c}{M_s} \right)^{5/7}.
\label{eq:mc-formula}
\end{equation}
Using $M_u \simeq 2 \times 10^{16}~{\rm GeV}$ and $M_s/M'_c \simeq 
16\pi^3/g^2 C \simeq 200$, we obtain $M'_c \simeq 5 \times 
10^{14}~{\rm GeV}$ and $M_s \simeq 1 \times 10^{17}~{\rm GeV}$.
These values become important when we discuss gauge and Yukawa 
coupling unifications in section \ref{sec:unif}.

The second important consequence of the large dimension is that 
it explains part of the observed structure of fermion masses. 
Yukawa interactions are forbidden by 5D supersymmetry 
from appearing in the bulk Lagrangian, and hence must be brane 
localized.  They are located on the $y=0$ brane 
\begin{equation}
  S = \int d^4x \; dy \; \delta(y)
    \biggl[ \int d^2\theta \Bigl( y_T \hat{T} \hat{T} H 
	+ y_F \hat{T} \hat{F} \bar{H} \Bigr) + {\rm h.c.} \biggr],
\label{eq:yukawa}
\end{equation}
where $\hat{T}$ ($\hat{F}$) runs over all the matter chiral superfields 
in the ${\bf 10}$ ($\bar{\bf 5}$) representation: brane-localized $T$ 
($F$) and bulk $T$ and $T'$ ($F$ and $F'$).  Since the full $SU(5)$ 
symmetry is operative at $y=0$, these Yukawa couplings must respect 
the $SU(5)$ symmetry.  This means that, if quarks and leptons are 
located on the brane, they respect $SU(5)$ mass relations.  The 
resulting 4D Yukawa couplings are suppressed by a factor of 
$1/(M_s R)^{1/2} \approx 0.1$ as the Higgs wavefunctions are 
spread out over the bulk.  On the other hand, if quarks and leptons 
are in the bulk, $u,e$ and $q$ ($d$ and $l$) arise from different 
hypermultiplets ${\cal H}_T$ and ${\cal H}_{T'}$ (${\cal H}_F$ and 
${\cal H}_{F'}$).  Therefore, they do not respect $SU(5)$ mass 
relations because the down-type quark and charged lepton masses 
come from different couplings, which are not related by the $SU(5)$ 
symmetry.\footnote{
We could also introduce Yukawa couplings for bulk matter on the 
$y=\pi R$ brane, which do not respect the $SU(5)$ symmetry.}
Moreover, since the matter wavefunctions are also spread out in the 
extra dimension, the resulting 4D Yukawa couplings receive a stronger 
suppression, by a factor of $1/(M_s R)^{3/2} \approx 10^{-3}$, than 
in the case of brane matter.  Thus we find a clearly successful 
correlation between the mass of the fermion and whether it has 
$SU(5)$ mass relations --- heavier fermions display $SU(5)$ mass 
relations while lighter ones do not.  Obviously, Yukawa couplings 
involving both bulk and brane matter receive a suppression factor 
of $1/(M_s R)$.

The location of some matter is determined because the extra dimension 
is ``large''.  Since our theory has $M_c \approx 10^{15}~{\rm GeV}$, 
the $X$ gauge bosons are considerably lighter, of mass about 
$10^{15}~{\rm GeV}$, than in the case of 4D supersymmetric grand 
unification. This makes dimension six proton decay a non-trivial issue 
in our theory; for instance, if all the matter fields were localized on 
the brane, the $X$ gauge boson exchange would induce proton decay at too 
rapid a rate.  We find that this rapid proton decay is avoided if the 
quarks and leptons of the first generation coming from a ${\bf 10}$ 
representation are bulk fields, since then $q$ and $u,e$ come from 
different hypermultiplets and the broken gauge boson exchange does not 
lead to proton decay.  We will say that $T_1$ is in the bulk, although 
we really mean the combination $\{ T_1,T^c_1 \} + \{ T'_1,T'^c_1 \}$.
On the other hand, the top quark must arise from a brane field $T_3$.
If the top quark were a bulk mode, it would have a mass suppressed 
by a factor of $1/(M_s R)^{3/2}$ giving too light a top quark, 
even in the case that the Yukawa interaction is strong at $M_s$. 
With $T_3$ on the brane, strong coupling leads to a top Yukawa 
coupling of the low energy theory of $4 \pi /(M_s R)^{1/2} \approx 1$, 
giving a top quark mass of the observed size.  Thus, given the existence 
of the large dimension of size $M_s R = O(100)$, we are able to derive 
the location of both the first and third generation ${\bf 10}$'s, and 
we find that at least some aspects of flavor physics are associated 
with the geometry of the extra dimension, and with strong coupling.  
Further consequences of the large size of the fifth dimension, for 
example for gaugino mass relations and Yukawa coupling unification, 
are discussed in later sections.

We now proceed further with matter geography by considering 
fermion mass relations.  The location of $F_3$ determines whether 
we have $b/\tau$ Yukawa unification, which gives a successful 
prediction for $m_b/m_\tau$ at $O(10\%)$ level in supersymmetric 
grand unified theories \cite{Chanowitz:1977ye}.  For most of this 
paper we choose to put $F_3$ on the $y=0$ brane to preserve $b/\tau$ 
unification. (The theory without $b/\tau$ unification is discussed in 
section \ref{sec:alt}.)  On the other hand, since the $SU(5)$ mass 
relation for $s/\mu$ does not work, either $T_2$ or $F_2$ must be 
located in the bulk.  Summarizing, we have derived the locations 
of $T_1$, $T_3$, and $F_3$ by considering dimension six proton decay, 
the size of the top Yukawa coupling, and $b/\tau$ Yukawa unification.  
We have also found that either $T_2$ or $F_2$ must be in the 
bulk to avoid an unwanted $SU(5)$ mass relation for $s/\mu$.
Therefore, we are left with $2^3-1 = 7$ possibilities for the matter 
location at this stage, corresponding to choices for the locations of 
$T_2$, $F_2$, and $F_1$.  Further determination must await the 
introduction of supersymmetry breaking in the next sub-section.

\subsubsection{$U(1)_R$ symmetry}

We here discuss the important issue of what further brane-localized 
operators can be introduced in the theory.  The 5D restricted gauge 
symmetry alone allows many unwanted operators on the branes.  For 
instance, the operators $[H \bar{H}]_{\theta^2}$ and $[F H]_{\theta^2}$ 
give a large mass, of order the unified scale, for the Higgs doublets 
destroying the solution to the doublet-triplet splitting problem, 
$[T F F]_{\theta^2}$ causes disastrous dimension four proton decay, and 
$[T T T F]_{\theta^2}$ induces too rapid dimension five proton decay. 
In addition, if matter is located in the bulk, $SU(5)$ 
non-invariant operators on the $y=\pi R$ brane, such as 
$[T_Q T_Q \bar{H}^c_T]_{\theta^2}$ and $[T_Q F_L H^c_T]_{\theta^2}$, 
reintroduce the problem of dimension five proton decay mediated by 
colored Higgsino exchange.  Remarkably, however, the structure of the 
theory allows a mechanism that simultaneously suppresses all of these 
unwanted operators \cite{Hall:2001pg}.  Since the bulk Lagrangian has 
higher dimensional supersymmetry, it possesses an $SU(2)_R$ symmetry.
It also has an $SU(2)_H$ flavor symmetry rotating the two Higgs 
hypermultiplets in the bulk.  After orbifolding, these two $SU(2)$ 
symmetries are broken to two $U(1)$ symmetries, one from 
$SU(2)_R$ and one from $SU(2)_H$.  A particularly interesting 
symmetry is the diagonal subgroup of these $U(1)$ symmetries, which 
we call $U(1)_R$ symmetry since it is an $R$ symmetry rotating the 
Grassmann coordinate of the low energy 4D $N=1$ supersymmetry.
We can extend this bulk $U(1)_R$ symmetry to the full theory 
by assigning appropriate charges to the brane-localized 
quark and lepton superfields, and use it to constrain possible forms 
of brane-localized operators.  The resulting $U(1)_R$ charges are 
given in Table~\ref{table:U1R}, where $T$ and $F$ represent both brane 
and bulk matter (primed fields have the same charges as unprimed fields).
Imposing this $U(1)_R$ symmetry on the theory, we can forbid unwanted 
operators while keeping the Yukawa couplings.  Proton decay from 
operators of dimension four and five are prohibited, and all 
$R$-parity violating operators are absent since $U(1)_R$ contains 
the usual $R$ parity as a discrete subgroup.  The $U(1)_R$ 
symmetry also forbids a bulk mass term for the Higgs hypermultiplets, 
$[H \bar{H} - H^c \bar{H^c}]_{\theta^2}$, which would remove 
the Higgs doublets from the low energy theory and reintroduce 
dimension five proton decay from colored Higgsino exchange.
Therefore, the $U(1)_R$ symmetry provides a complete solution to 
the doublet-triplet and proton decay problems.
\begin{table}
\begin{center}
\begin{tabular}{|c|cc|cccc|cccc|}  \hline 
  & $V$ & $\Sigma$ & $H$ & $H^c$ & $\bar{H}$ & $\bar{H}^c$ 
  & $T$ & $T^c$ & $F$ & $F^c$ \\ \hline
  $U(1)_R$ & 0 & 0 & 0 & 2 & 0 & 2 & 1 & 1 & 1 & 1 \\ \hline
\end{tabular}
\end{center}
\caption{$U(1)_R$ charges for 4D vector and chiral superfields.}
\label{table:U1R}
\end{table}

The above $U(1)_R$ is broken to its $R$-parity subgroup by 
supersymmetry breaking, introduced in the next sub-section.  
Since the breaking scale is small, however, it will not reintroduce 
the problem of proton decay.  The presence of an $R$ symmetry broken 
only through supersymmetry breaking effects is also important for 
generating the supersymmetric mass term for the two Higgs doublets 
(the $\mu$ term) at the correct size of the order of the weak 
scale \cite{Hall:2002up}.

\subsection{Supersymmetry breaking from boundary conditions}
\label{subsec:theory2}

Having obtained a unified theory free from the problems of usual 
supersymmetric grand unification, we now introduce supersymmetry 
breaking into the theory.  There are two natural ways of introducing 
supersymmetry breaking in theories with unified scale extra dimensions: 
one is through a supersymmetry breaking expectation value of some 
brane-localized 4D field \cite{Randall:1998uk, Kaplan:1999ac} and the 
other is through boundary conditions \cite{Barbieri:2001yz}.  In the 
former case, there is a 4D chiral superfield $Z$ whose highest component, 
$F_Z$, has a non-vanishing vacuum expectation value.  However, in 
our theory some of the lightest two generations of matter propagate 
in the bulk, so that this way of introducing supersymmetry breaking 
leads to the supersymmetric flavor problem through direct couplings 
such as $[Z^\dagger Z\, T_i^\dagger T_j]_{\theta^2 \bar{\theta}^2}$.
Therefore, we choose the latter case where supersymmetry is broken 
by boundary conditions. 

One important feature of boundary condition breaking is that all of 
the supersymmetry breaking parameters are completely specified in 
terms of a single continuous parameter $\alpha$ \cite{Barbieri:2001dm}.
The supersymmetry breaking is introduced by imposing the boundary 
conditions such that, under the orbifold translation 
${\cal T}: y \rightarrow y + 2\pi R$, the component fields are rotated 
by a $U(1)$ subgroup of $SU(2)_R$ which does not commute with the 
orbifold reflection ${\cal Z}$; the angle of the rotation is 
parameterized by $\alpha$.  Since $A_M$ and $\sigma$ are singlet 
under $SU(2)_R$, only the two gauginos, $\lambda$ and $\lambda'$, 
are subject to the above rotation in the gauge multiplet.  
Similarly, for hypermultiplets, only scalar components, $\phi$ and 
$\phi^c$, receive the rotation.  Then, without a loss of generality, 
we can take the boundary conditions for the gauginos and scalar 
components of hypermultiplets as 
\begin{equation}
  \pmatrix{\lambda^{(p)} \cr \lambda'^{(p)}}(x^\mu,y) 
  = \pmatrix{\lambda^{(p)} \cr -\lambda'^{(p)}}(x^\mu,-y) 
  = e^{2\pi i \alpha \sigma_2} p
    \pmatrix{\lambda^{(p)} \cr \lambda'^{(p)}}(x^\mu,y+2\pi R),
\label{eq:bc-lambda}
\end{equation}
and 
\begin{equation}
  \pmatrix{\phi^{(p)} \cr \phi^{c(p)\dagger}}(x^\mu,y) 
  = \pmatrix{\phi^{(p)} \cr -\phi^{c(p)\dagger}}(x^\mu,-y) 
  = e^{2\pi i \alpha \sigma_2} p\, \eta_{\Phi}
    \pmatrix{\phi^{(p)} \cr \phi^{c(p)\dagger}}(x^\mu,y+2\pi R), 
\label{eq:bc-phi}
\end{equation}
respectively, where $\sigma_{1,2,3}$ are the Pauli spin matrices.
All the other component fields obey the same boundary conditions 
as before: Eqs.~(\ref{eq:bc-g}, \ref{eq:bc-h}).  

After KK decomposition, the above boundary conditions generate soft 
supersymmetry breaking masses of order $\alpha/R$. 
Since $1/R \approx 10^{15}~{\rm GeV}$, $\alpha$ must be an extremely 
small number, $\alpha \approx 10^{-13}$.  However, this does not mean 
that we need a fine tuning to obtain supersymmetry breaking masses 
of the order of the weak scale.  In fact, it has been shown 
that the above supersymmetry breaking twist by $\alpha$ is equivalent 
to having a supersymmetry breaking vacuum expectation value for 
an auxiliary field in the 5D gravity multiplet \cite{Marti:2001iw}.
In other words, by making a suitable gauge transformation that depends 
on the coordinate $y$, we can always go to the basis where the boundary 
conditions do not have any supersymmetry breaking twist and all the 
supersymmetry breaking effects are contained in the vacuum expectation 
value of some auxiliary field.  Since a vacuum expectation value of 
the auxiliary field can be generated dynamically through strongly 
coupled gauge interactions, having a small $\alpha$ parameter in the 
original basis is completely natural in this case.  Note that this 
situation is quite different from the case of the $SU(5)$ breaking 
in the previous section, where the boundary conditions do not contain 
any continuous parameter and the breaking cannot be viewed as a 
``spontaneous breaking'' that arises entirely from an expectation 
value of some background field \cite{Hall:2001tn}.

We now explicitly calculate the soft supersymmetry breaking terms 
resulting from the boundary conditions Eqs.~(\ref{eq:bc-lambda}, 
\ref{eq:bc-phi}).  Below the compactification scale, the theory is 
reduced to the usual 4D MSSM.  Under the KK decomposition, the MSSM 
gauginos $\lambda^a$ ($a=1,2,3$) are contained in the two 5D gaugino 
fields, $\lambda$ and $\lambda'$, as
\begin{equation}
  \pmatrix{\lambda \cr \lambda'}(x^\mu,y) 
  = \pmatrix{\lambda^a(x^\mu)\cos(\alpha y/R) \cr 
    \lambda^a(x^\mu)\sin(\alpha y/R)} + \cdots.
\label{eq:gaugino-0}
\end{equation}
The MSSM Higgs fields $h$ arise from the two scalar fields, 
$\phi$ and $\phi^c$, in the corresponding hypermultiplets as
\begin{equation}
  \pmatrix{\phi \cr \phi^{c\dagger}}(x^\mu,y) 
  = \pmatrix{h(x^\mu)\cos(\alpha y/R) \cr 
    h(x^\mu)\sin(\alpha y/R)} + \cdots,
\label{eq:higgs-0}
\end{equation}
and similarly for the squarks and sleptons located in the bulk.  
The Higgsinos and the bulk quarks and leptons are zero modes of the 
$\psi$ fields in the corresponding hypermultiplets, and the gauge 
bosons come from zero modes of $A_\mu$. The supersymmetry breaking 
terms are then obtained by substituting these KK mode expansions 
into the original 5D action.

What supersymmetry breaking operators do we obtain?  To answer this 
question, we first consider only the bulk interactions (kinetic terms).
In our theory all the supersymmetry breaking effects are encoded in the 
KK mode decompositions Eqs.~(\ref{eq:gaugino-0}, \ref{eq:higgs-0}), 
which have a $y$-dependent twist by the $U(1)$ subgroup of $SU(2)_R$. 
This implies that supersymmetry breaking operators arise only from 
the terms which contain the $y$ derivative, since the bulk Lagrangian 
possesses a global $SU(2)_R$ symmetry. The resulting supersymmetry 
breaking interactions are soft by dimensional analysis: the derivative 
$\partial_y$ becomes the supersymmetry breaking parameter $\alpha/R$, 
which has a positive mass dimension.  This argument can be easily 
extended to the case with brane-localized interactions.  The 
brane-localized interactions can always be made invariant under the 
$U(1) \subset SU(2)_R$; to be more precise, we can always choose brane 
interactions at $y=2\pi n R$ ($n = \pm 1, \pm 2, \cdots$) so that the 
whole system is invariant under the $U(1)$ rotation by $\alpha$ and 
the spacetime translation $y \rightarrow y + 2\pi R$.  Then, we can 
show that all the supersymmetry breaking terms still arise only from 
the terms which contain the derivative $\partial_y$.

The most obvious place where the $y$ derivative appears is the 5D 
kinetic terms.  For the gauginos, the 5D kinetic term contains the 
term $-\lambda \partial_y \lambda' + {\rm h.c.}$, which gives Majorana 
mass terms for the MSSM gauginos, $-(\alpha/2R) \lambda^a \lambda^a 
+ {\rm h.c.}$, after integrating over $y$.  The Higgs bosons obtain 
the soft supersymmetry breaking mass terms, $-(\alpha/R)^2 h^\dagger h$, 
from the term $-\partial_y \phi^\dagger \partial_y \phi - \partial_y 
\phi^{c\dagger} \partial_y \phi^c$ in the 5D kinetic terms for the 
scalar fields.  The same soft supersymmetry breaking mass is 
also induced for squarks and sleptons living in the bulk.
In addition to these obvious contributions, the presence of the 
brane-localized Yukawa couplings also provides a somewhat non-trivial 
source for soft supersymmetry breaking parameters.  Suppose we have a 
superpotential interaction $[\kappa \Phi_1 \Phi_2 \Phi_3]_{\theta^2}$ 
on the $y=0$ brane.  Then, if $\Phi_3$ is a bulk field, we obtain the 
interaction term $\delta(y)\kappa\phi_1\phi_2\partial_y\phi_3^{c\dagger} 
+ {\rm h.c.}$ in the Lagrangian.  (In the 4D superfield language, 
this term arises from eliminating the auxiliary field of $\Phi_3$, 
since the bulk kinetic term contains a superpotential term 
$[\Phi_3 \partial_y \Phi_3^c]_{\theta^2}$ \cite{Arkani-Hamed:2001pv}.)
This term provides a trilinear scalar interaction for low energy 
modes $\phi_{i,0}$ with the coefficient $\kappa\alpha/R$: 
$(\kappa\alpha/R)\phi_{1,0}\phi_{2,0}\phi_{3,0} + {\rm h.c.}$.
Since exactly the same contribution is obtained from the term 
$\delta(y)\kappa\phi_2\phi_3\partial_y\phi_1^{c\dagger}$ 
($\delta(y)\kappa\phi_3\phi_1\partial_y\phi_2^{c\dagger}$) if $\Phi_1$ 
($\Phi_2$) is in the bulk, we finally obtain the trilinear scalar 
interaction, $\phi_{1,0}\phi_{2,0}\phi_{3,0} + {\rm h.c.}$, with a
coefficient given by $\kappa\alpha/R$ times the number of $\phi_i$ 
fields propagating in the bulk.  Clearly, no soft supersymmetry 
breaking masses are generated for brane-localized fields.

To summarize, the soft supersymmetry breaking terms in our theory 
are given by
\begin{eqnarray}
  {\cal L}_{\rm soft} &=& 
    - \frac{1}{2} \left( \tilde{m} \lambda^a \lambda^a + {\rm h.c.} \right)
    - \tilde{m}^2 h^\dagger h 
    - \tilde{m}^2 \tilde{f}_B^\dagger \tilde{f}_B 
\nonumber\\
 && + \left( y_f \tilde{m} \tilde{f}_b \tilde{f}_b h 
        + 2 y_f \tilde{m} \tilde{f}_B \tilde{f}_b h
        + 3 y_f \tilde{m} \tilde{f}_B \tilde{f}_B h + {\rm h.c.} \right),
\label{eq:soft}
\end{eqnarray}
where $h$, $\tilde{f}_B$ and $\tilde{f}_b$ collectively represent the 
two Higgs doublets, squarks/sleptons in the bulk and squarks/sleptons
on the brane, respectively.  Here, we have defined 
$\tilde{m} \equiv \alpha/R$, and $y_f$ is the value of the 
corresponding Yukawa coupling. Since supersymmetry breaking effects 
from boundary conditions are exponentially shut off above the 
compactification scale, the soft supersymmetry breaking masses in 
Eq.~(\ref{eq:soft}) must be regarded as the running mass parameters 
at the compactification scale $M'_c$.  We also explicitly see that 
the supersymmetry breaking terms in Eq.~(\ref{eq:soft}) preserve 
the discrete $R$-parity subgroup of the $U(1)_R$ symmetry given in 
Table~\ref{table:U1R}.  This is because the boundary conditions 
Eqs.~(\ref{eq:bc-lambda}, \ref{eq:bc-phi}) always rotate two fields 
which differ by two units in their $U(1)_R$ charges. (In another basis, 
it comes from the fact that the auxiliary field having a supersymmetry 
breaking vacuum expectation value has a $U(1)_R$ charge of $-2$.)
Thus, we have $R$-parity conservation in our theory, and the lightest 
supersymmetric particle (LSP) is completely stable.

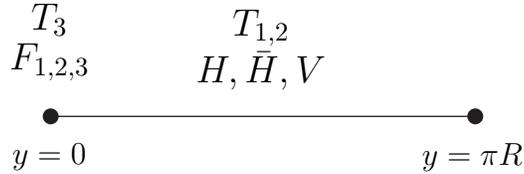
\begin{figure}
\begin{center} 
\begin{picture}(100,70)(150,20)
  \Line(120,50)(280,50)
  \Vertex(120,50){3} \Text(120,40)[t]{$y=0$}
  \Vertex(280,50){3} \Text(280,40)[t]{$y=\pi R$}
  \Text(120,83)[b]{\large $T_3$} \Text(120,68)[b]{\large $F_{1,2,3}$}
  \Text(200,81)[b]{\large $T_{1,2}$} 
  \Text(200,63)[b]{\large $H, \bar{H}, V$}
\end{picture}
\caption{Locations of $SU(5)$ matter, Higgs and gauge multiplets 
 in the fifth dimension.}
\label{fig:theory}
\end{center}
\end{figure}

Now, we consider consequences of the supersymmetry breaking terms 
given in Eq.~(\ref{eq:soft}).  We first note that squarks and sleptons 
in the bulk have non-vanishing soft masses at the compactification 
scale while those on the brane do not.  Therefore, the brane and bulk 
scalars have different masses at the weak scale, even if they have 
the same standard model quantum numbers.  This implies that the first 
two generation fields having the same gauge quantum number must be 
located in the same place to evade stringent constraints from flavor 
changing neutral current processes.  This consideration immediately 
fixes the location of $T_2$ to be the bulk, since $T_1$ must be the 
bulk field to evade constraints from proton decay; $F_1$ and $F_2$ 
must also be put together either in the bulk or on the $y=0$ brane.  
Thus, we are finally left with only two choices for the matter location: 
whether we put $F_{1,2}$ in the bulk or on the brane.  The locations 
for the other matter fields are completely fixed: $T_3$ and $F_3$ on 
the brane and $T_{1,2}$ in the bulk.  As we will see later, the case 
of $F_{1,2}$ in the bulk leads to large lepton flavor violating 
processes, which pushes up the overall scale for supersymmetry breaking 
masses and leads to a fine tuning for electroweak symmetry breaking.
Therefore, if we want to keep all desired features including $b/\tau$ 
unification and naturalness for electroweak symmetry breaking, we end up 
with a unique possibility for the location of matter.  The resulting 
theory is illustrated in Fig.~\ref{fig:theory}.  The case without 
$b/\tau$ unification will be discussed in section \ref{sec:alt}.

We here comment on the corrections to the soft supersymmetry breaking 
parameters at the compactification scale $M'_c$.  The soft parameters 
in Eq.~(\ref{eq:soft}) were derived at tree level by considering the 
bulk kinetic terms and the brane-localized Yukawa couplings. 
There are two possible sources for the corrections to these values. 
One comes from brane-localized kinetic terms for the bulk fields, and 
the other from finite loop radiative corrections.  We first consider 
the effect from brane-localized kinetic terms.  If there were no brane 
kinetic terms, the gaugino masses would be unified at $M'_c$, giving 
gaugino mass ratios at low energies, $m_i/m_j$, different from 
that in the case of conventional one-scale unification; 
the difference amounts to more than $20\%$ for $m_3/m_1$, leading 
to observable consequences.  However, this does not happen 
due to the presence of brane-localized kinetic operators at $M'_c$. 
After KK decomposition, the quadratic part of the MSSM gaugino 
Lagrangian is
\begin{equation}
  {\cal L} = {1 \over g_a^2} 
    \lambda^{a\dagger} i \bar{\sigma}^\mu \partial_\mu \lambda^a
    - {1 \over 2} {\tilde{m} \over g_*^2} \lambda^a \lambda^a
\label{eq:gaugino}
\end{equation}
where $1/g_a^2 = 1/g_*^2 + 1/\tilde{g}_a^2$, 
$1/g_*^2 \equiv \pi R/g_5^2$, and $1/\tilde{g}_a^2$ ($\ll 1/g_*^2$) 
represent the sum of the two contributions from brane gauge kinetic 
terms at $y=0$ and $\pi R$.  Here, all the quantities are evaluated 
at the compactification scale, and $g_a$ are the 4D gauge couplings 
at $M'_c$ with effects from logarithmic running above $M'_c$ included 
via $\tilde{g}_a$.  Rescaling the gaugino fields to canonical 
normalization leads to the gaugino masses $m_a = (g_a^2/g_*^2)\tilde{m}$ 
at $M'_c$, and thus $m_a/g_a^2$ are universal.  Therefore, even though 
gaugino masses are generated at a smaller scale $M'_c$ than the gauge 
coupling unification scale, we find that the grand unified relation 
for the gaugino masses, $m_1/g_1^2 = m_2/g_2^2 = m_3/g_3^2$, holds 
very precisely.

How about brane-localized kinetic terms for hypermultiplets?  
Since brane kinetic terms contribute to the 4D kinetic terms after 
the KK decomposition, they modify the soft masses through 
wavefunction renormalization.  However, since effects from 
brane-localized terms are generically suppressed by the volume factor 
$M'_c/M_s$, they only give corrections to the soft masses of order 
$(M'_c/M_s)\tilde{m}^2 \approx 10^{-2}\tilde{m}^2$. Although these 
contributions are flavor non-universal, this amount of flavor 
violation does not contradict with flavor changing neutral current 
experiments \cite{Gabbiani:1996hi}; the constraints from leptonic 
processes, such as $\mu \rightarrow e\gamma$, are evaded for 
$\tilde{m} \gsim 200~{\rm GeV}$, and the bounds from hadronic ones, 
such as $K$-$\bar{K}$ mixing, are also avoided in the same parameter 
region partly because in the down sector only the left-handed squarks 
receive a flavor non-universality from the boundary conditions 
and partly because the gluino adds a large universal contribution 
through the renormalization group evolution below $M'_c$. In principle, 
the logarithmic radiative corrections from an energy interval between 
$M_s$ and $M'_c$ could enhance the contribution from the brane kinetic 
terms by a factor of $\ln(M_s/M'_c)$.  However, we find from dimensional 
analysis that Yukawa loops always give power divergences to the brane 
kinetic terms and thus do not enhance the contributions. The gauge 
loops could provide the enhancements, but they are flavor universal 
and thus have little observable consequences.

As for finite loop radiative corrections, they arise from non-local 
operators spread out in the extra dimension and appear as threshold 
effects at $M'_c$ in a 5D calculation. In the 4D picture this 
corresponds to including supersymmetry breaking effects from higher 
KK modes. These contributions are shut off above $M'_c$ and give only 
one-loop suppressed corrections, of order $1/16\pi^2 \approx 10^{-2}$, 
to the soft mass parameters.  Thus these threshold contributions are 
smaller than the usual logarithmic contributions coming from 
renormalization group evolutions below $M'_c$, by a factor of 
$\ln(M'_c/M_Z)$, so that we can safely neglect them.  In fact, this 
approximation is well justified because all the MSSM couplings are 
sufficiently weak at the compactification scale.

So far, we have considered only the soft supersymmetry breaking 
parameters arising from the kinetic terms and the Yukawa couplings. 
However, to understand the low energy physics, we also have to specify 
the Higgs sector.  In particular, both the supersymmetric mass ($\mu$ 
parameter) and the holomorphic supersymmetry breaking mass ($\mu B$ 
parameter) for the two Higgs doublets must be of the order of the weak 
scale to obtain viable phenomenology. In our theory these parameters 
can be naturally generated in a number of ways, and we will discuss 
some explicit examples in sub-section \ref{subsec:mu}.  However, to 
keep the analysis as general as possible, here we treat $\mu$ and 
$\mu B$ as free parameters.  Then, all the supersymmetry breaking 
parameters in our theory are completely specified by the following 
four parameters: $M'_c$, $\tilde{m}$, $\mu$ and $\mu B$.  Among them, 
the last two parameters are related to the electroweak vacuum 
expectation value, $v \equiv (\vev{h_u}^2 + \vev{h_d}^2)^{1/2}$, and 
the ratio of vacuum expectation values for the two Higgs doublets, 
$\tan\beta \equiv \vev{h_u}/\vev{h_d}$, through the conditions of 
electroweak symmetry breaking. Here $h_u$ and $h_d$ are the 
two Higgs doublets of the MSSM giving up-type and down-type quark 
masses, respectively.  Since we know that $M'_c \approx 10^{15}~{\rm GeV}$ 
and $v \simeq 175~{\rm GeV}$, we are finally left with only two free 
parameters, $\tilde{m}$ and $\tan\beta$, to specify the superparticle 
spectrum. In general there is also a phase for $\mu$ which is not 
determined by the condition of electroweak symmetry breaking; however, 
if this phase is far from $\pm 1$, it will lead to excessively large 
contributions to electric dipole moments, and hence we allow only 
this discrete choice.  In sub-section \ref{subsec:mu} we give a natural 
origin for $\mu$ in our theory, and find that it is indeed real, 
solving the supersymmetric $CP$ problem.  Note that various low energy 
quantities, including supersymmetric ones, are calculable in terms of 
these two parameters (and the sign of $\mu$) and the ratio of the 
cutoff and compactification scales, $M_s/M'_c$.

Now, we present the result for the superparticle spectrum at the 
weak scale in our theory.  The soft supersymmetry breaking parameters 
at $M_Z$ are obtained by evolving the boundary values at $M'_c$, given 
in Eq.~(\ref{eq:soft}), using renormalization group equations.
In Table~\ref{table:soft}, results for the soft mass parameters 
at $M_Z$ are given, in GeV, for the three MSSM gauginos, $\lambda^a$, 
for the up-type (down-type) Higgs bosons, $h_u$ ($h_d$), for the first 
two generation squarks and sleptons living in the bulk (on the brane), 
$\tilde{f}_B$ ($\tilde{f}_b$), and for the third generation squarks and 
sleptons, $\tilde{f}_3$.  The parameters $A_t$, $A_b$ and $A_\tau$ are 
the trilinear couplings for squarks and sleptons of the third generation 
defined by ${\cal L} = - y_f A_f \tilde{f} \tilde{f} h + {\rm h.c.}$.
We have given the soft masses for the first two generations of squarks 
and sleptons in the cases of both bulk and brane locations to maintain 
some generality, although these locations are completely fixed in 
the present model, giving $\tilde{q}_{1,2} = \tilde{q}_B$, 
$\tilde{u}_{1,2} = \tilde{u}_B$, $\tilde{d}_{1,2} = \tilde{d}_b$, 
$\tilde{l}_{1,2} = \tilde{l}_b$, and $\tilde{e}_{1,2} = \tilde{e}_B$.

In the table, we have taken the two free parameters to be $\tilde{m} = 
200~{\rm GeV}$ and $\tan\beta = 5$ as a representative value. However, 
the dependence on $\tilde{m}$ is quite simple so that we can easily 
read off the soft masses for any value of $\tilde{m}$: all the numbers 
scale almost linearly with $\tilde{m}$.  The dependence on $\tan\beta$ is 
somewhat more complicated, but for moderately small values for $\tan\beta$ 
($\tan\beta \lsim 10$), the resulting soft masses are almost 
insensitive to the value of $\tan\beta$, except that $|B|$ is almost 
proportional to $1/\tan\beta$.  When $\tan\beta$ is further increased, 
several quantities vary because the bottom and tau Yukawa couplings 
become large.  In particular, the soft mass squared for the 
right-handed stau becomes negative for $\tan\beta \gsim 25$, so that 
our theory does not allow very large values for $\tan\beta$.
\begin{table}
\begin{center}
\begin{tabular}{|cc|cc|cc|cc|cc|}  \hline 
  $\lambda^3$ & $480$ & $\tilde{q}_B$ & $480$ & $\tilde{q}_b$ & $440$ 
    & $\tilde{q}_3$ & $390$ & $A_t$    & $-410$ \\
  $\lambda^2$ & $170$ & $\tilde{u}_B$ & $470$ & $\tilde{u}_b$ & $420$ 
    & $\tilde{u}_3$ & $310$ & $A_b$    & $-730$ \\
  $\lambda^1$ & $85$  & $\tilde{d}_B$ & $470$ & $\tilde{d}_b$ & $420$ 
    & $\tilde{d}_3$ & $420$ & $A_\tau$ & $-320$ \\ 
    \cline{1-2} \cline{9-10}
  $h_u$    & $280\,i$ & $\tilde{l}_B$ & $240$ & $\tilde{l}_b$ & $140$ 
    & $\tilde{l}_3$ & $140$ & $|\mu|$  & $280$  \\
  $h_d$    & $240$    & $\tilde{e}_B$ & $210$ & $\tilde{e}_b$ & $67$ 
    & $\tilde{e}_3$ & $66$  & $|B|$    & $95$   \\ \hline
\end{tabular}
\end{center}
\caption{The soft supersymmetry breaking parameters in GeV for 
 $\tilde{m} = 200~{\rm GeV}$ and $\tan\beta = 5$.}
\label{table:soft}
\end{table}

We finally comment on several phenomenological features of the 
spectrum.  In the MSSM the tree-level mass for the lightest $CP$ 
even Higgs boson is smaller than $M_Z \cos(2\beta)$, so that the 
experimental limit requires a sizable radiative contribution. 
This can occur from top loop corrections to the Higgs quartic 
interaction \cite{Okada:1990vk}, but, in several schemes for 
supersymmetry breaking, this requires the top squark to be very
heavy, increasing the amount of fine tuning for successful electroweak
symmetry breaking. This is not the situation for our boundary 
condition supersymmetry breaking because of the large $A$ parameter 
generated for the top quark; the predicted sign of $A$ in 
Eq.~(\ref{eq:soft}) is such that, on scaling to the infrared, 
$|A|$ is increased by the radiative contribution from the 
gaugino mass.  As a result, the left-right mixing of the top squarks 
is large, which increases the radiative contribution to the Higgs mass. 
We find that for $\tan \beta = 3$ $(5)$ the Higgs mass obeys the 
experimental bound for $\tilde{m} \gsim 400$ $(200)$ GeV. 

Of the superpartners listed in Table~\ref{table:soft}, two have mass 
parameters which are significantly smaller than the rest: the 
right-handed scalar tau and the bino. The experimental collider 
signatures for our theory depend crucially on which of these particles 
is lighter, and yet, for several reasons, this cannot be definitively 
predicted.  Typically we find the scalar tau to be lighter, but there 
are effects which could reverse the order.  First, there are 
uncertainties in the mass parameters listed in Table~\ref{table:soft}. 
To construct a complete theory, with a natural origin for 
both neutrino masses and the $\mu$ parameter, in section 
\ref{sec:develop} we introduce an additional gauge interaction, 
$U(1)_X$ ($\subset SO(10)/SU(5)$), broken at some high energy scale 
below $M'_c$ to generate Majorana masses for the right-handed 
neutrinos.  In that case, the $U(1)_X$ gaugino mass leads to an 
additional radiative contribution to masses of the scalar superpartners 
which depends on the $U(1)_X$ breaking scale. This will give only 
a small percentage correction for most of the superpartners, but for 
the right-handed stau the correction could be sizable because it has 
a small mass. This alone may make the stau heavier than the bino 
for large values of the $U(1)_X$ gauge coupling. 
Second, there are additional effects in going from the mass parameters 
of Table~\ref{table:soft} to the physical particle masses. All scalar 
superpartners receive corrections from electroweak symmetry breaking 
via the electroweak $D$ terms. Again this is most important for the 
scalar tau, raising its mass from $66~{\rm GeV}$ to $77~{\rm GeV}$ 
for $\tilde{m} = 200~{\rm GeV}$. Hence we discuss collider 
phenomenology for both cases of stau and bino ``LSP''.\footnote{
Our theory possesses a continuous $U(1)_R$ symmetry which is 
spontaneously broken to $R$ parity by the dynamics for an auxiliary 
field in the 5D supergravity multiplet, as discussed in sub-section 
\ref{subsec:r-axion}, so that the true LSP may be the $R$ axino.}
In either case some care is required in obtaining a precise value for 
the lower bound on $\tilde{m}$ from experiment, but we typically expect 
this bound to be in the region of $200~{\rm GeV}$.

If the stau is lighter than the bino it will appear as a charged
stable particle in collider detectors. The present limit on the 
mass of such a stau from the LEP experiments is about 
$100~{\rm GeV}$, coming from direct Drell-Yan production of 
$\tilde{\tau}^+ \tilde{\tau}^-$. At hadron colliders, in addition to 
Drell-Yan production, stau pairs arise from squark and gluino pair 
production followed by cascade decays, giving equal numbers of 
like sign and opposite sign events. If the bino is lighter, the 
signal events, arising from squark and gluino pair production, 
contain jets and missing transverse energy --- a classic signal for 
many supersymmetric theories. In this case the present experimental 
limit from LEP experiments on the stau mass is quite weak if it is 
close in mass to the bino.

Ultimately, an important experimental test of our theory is to 
measure the superpartner masses with sufficient accuracy to probe 
the location of the matter: at the compactification scale, all bulk 
superpartners have a universal mass, while all brane superpartners 
are massless.  In the physical spectrum this is clearly manifested 
in the lepton sector, with $\tilde{l}$ and $\tilde{e}_3$ much 
lighter than $\tilde{e}_{1,2}$, but is harder to uncover in the 
squarks due to the gluino focusing effect.

\section{Consequences}
\label{sec:unif}

In the previous section we have explicitly demonstrated how to break
both $SU(5)$ gauge symmetry and 4D supersymmetry by boundary conditions 
that act in the same spatial dimension. This leads to a highly 
constrained theory with many consequences, some of which we explore 
in this section. The gross flavor structure of the theory is 
determined by the location of matter in the fifth dimension, and 
in the next sub-section we show how many realistic features of the 
quark and lepton mass matrices emerge automatically. (In the next
section we explore the superpartner mass matrices, and give
predictions for a variety of flavor changing phenomena.)
The superpartner spectrum is so tightly constrained that we are able 
to evaluate the supersymmetric threshold corrections to the predictions 
to gauge and Yukawa coupling unification. Corrections to Yukawa 
unification from the compactification scale are also computed. 
Finally we study proton decay and the consequences of spontaneously 
breaking $U(1)_R$ symmetry.

\subsection{Quark and lepton mass matrices}
\label{subsec:masses}

With the matter configuration determined in the previous section, 
quark and lepton mass matrices take the form 
\begin{equation}
  {\cal L}_4 \approx 
  \pmatrix{
     T_1 & T_2 & T_3 \cr
  }
  \pmatrix{
     \epsilon^2 & \epsilon^2 & \epsilon      \cr
     \epsilon^2 & \epsilon^2 & \epsilon      \cr
     \epsilon   & \epsilon   & \underline{1} \cr
  }
  \pmatrix{
     T_1 \cr T_2 \cr T_3 \cr
  } H
\nonumber\\
  + \pmatrix{
     T_1 & T_2 & T_3 \cr
  }
  \pmatrix{
     \epsilon      & \epsilon      & \epsilon      \cr
     \epsilon      & \epsilon      & \epsilon      \cr
     \underline{1} & \underline{1} & \underline{1} \cr
  }
  \pmatrix{
     F_1 \cr F_2 \cr F_3 \cr
  } \bar{H}.
\label{eq:fermion-yukawa}
\end{equation}
Here we display only the gross structure that follows from the 
large size of the extra dimension, via the volume suppression factor 
$\epsilon \simeq (M'_c/M_s)^{1/2} \approx 0.1$, and suppress the 
coupling parameters of the brane-localized Yukawa interactions.  
Only underlined entries respect $SU(5)$, since the other entries 
involve $T_{1,2}$ which actually represent quarks and leptons from 
differing $SU(5)$ bulk multiplets.  The only unified mass relation 
is for $b/\tau$.

Although the matrices of Eq.~(\ref{eq:fermion-yukawa}) do not provide
a complete understanding of quark and charged lepton masses and 
mixings, they do capture many of the qualitative features. 
The masses of the first two generations are suppressed 
compared with the third one, by $\epsilon^2$ in the up quark 
sector and by $\epsilon$ in the down quark and charged lepton 
sectors. Furthermore, the quark mixing is small between the first 
two generations and the third. These are perhaps the most striking 
features of the data, and they result in our theory entirely
because $T_{1,2}$ alone reside in the bulk.

The most striking prediction of these matrices is that the flavor
mixing angles are of order unity for fields in $F_i$, which is
particularly important for the neutrinos. In sub-section 
\ref{subsec:neutrino} we discuss how neutrino masses are generated 
in our theory by the see-saw mechanism, leading to the operators 
$[LLHH]_{\theta^2}$ in the low energy theory. Because $F_i$ all 
reside at the same location, we predict that the coefficients of 
these operators have comparable sizes for all flavor combinations. 
Such an anarchy can reproduce observed neutrino phenomenology 
\cite{Hall:1999sn}, but is only consistent with the
large angle MSW interpretation of the solar neutrino flux.

The matrices of Eq.~(\ref{eq:fermion-yukawa}) leave two important
features of the spectrum unexplained: the $t/b$ ratio and the
hierarchy within the first two generations.  The first requires 
either $\tan\beta \approx 50$ or an additional suppression of 
$TF\bar{H}$ relative to $TTH$.  A very large value for $\tan\beta$ 
is not preferred, since the resulting large value for the tau Yukawa 
coupling leads to a scaling of $m_{\tilde{\tau}}^2$ to negative values 
at the weak scale, although this could be counteracted by a large 
contribution from the $U(1)_X$ gaugino radiative correction. Also, 
small values of $\tan\beta$ are excluded by the experimental limit 
on the mass of the lightest Higgs boson, unless the low energy 
theory is extended to include a singlet chiral superfield.  Thus 
we will typically use $\tan\beta \sim 5{\rm -}10$, giving a ratio of 
$b$ to $t$ Yukawa couplings of $\sim 1/5{\rm -}1/10$.

There are at least three ways to obtain a hierarchy within the light
two generations. One possibility is that there is a sufficient spread 
in the Yukawa coupling parameters to accommodate the data. While this 
is plausible for the Cabibbo angle, and perhaps also for the $d/s$ 
mass ratio, it seems less likely for the $u/c$ and $e/\mu$ mass 
ratios. A second possibility is to introduce a $U(1)$ flavor symmetry 
giving $T_1$ a different charge to $T_2$. For example, consider
the charge assignments $T_1(1)$, $T_{2,3}(0)$, $F_{1,2,3}(1)$, where
the charges for $F_{1,2,3}$ also give a suppression for the $b/t$ 
mass ratio. Taking the $U(1)$ symmetry breaking parameter 
$\delta \approx \epsilon$, we obtain a realistic pattern of masses 
$m_t:m_c:m_u \approx 1:\epsilon^2:\epsilon^4$, $m_b:m_s:m_d \approx 
m_\tau:m_\mu:m_e \approx 1:\epsilon:\epsilon^2$ and mixing angles 
$(V_{us},V_{cb},V_{ub}) \approx (\epsilon,\epsilon,\epsilon^2)$.  
A third possibility is to introduce an $S_2$ discrete symmetry which
interchanges $T_1 \leftrightarrow T_2$. This forces the couplings of
$T_1$ and $T_2$ to be equal, so that the second generation is $T^+ =
T_1 + T_2$, while the first generation $T^- = T_1 - T_2$ is massless
in the symmetry limit. The $u$ and $d$ masses and Cabibbo angle now
arise from small $S_2$ breaking couplings.

\subsection{Gauge coupling unification}
\label{subsec:gcu}

A theory with a supersymmetric desert and high scale gauge coupling 
unification leads to a prediction for the QCD coupling of the form
\begin{equation}
  \alpha_s(M_Z) = 0.1305 + \delta\alpha_s|_{\rm susy} 
                  + \delta\alpha_s|_{\rm u}.
\label{eq:alphasgen}
\end{equation}
The central number assumes that the $SU(5)$ split multiplets have 
the same form as in the MSSM \cite{Langacker:1995fk}, and should be 
compared with the experimental value \cite{Groom:2000in}
\begin{equation}
  \alpha_s^{\rm exp} = 0.117 \pm 0.002.
\label{eq:as-exp}
\end{equation}
In conventional supersymmetric grand unified theories, although the
prediction from gauge coupling unification is highly successful, 
the agreement with data is certainly not perfect.  Significant
threshold corrections from supersymmetric or unified thresholds are
required.  The weak scale threshold correction depends on the masses 
of the superpartners and Higgs bosons and has the form 
\begin{equation}
  \delta \alpha_s|_{\rm susy} 
    \simeq -0.0030 \sum_r {5 \over 38} 
    \left( 4b_1^r - \frac{48}{5}b_2^r + \frac{28}{5}b_3^r \right) 
    \ln\frac{m_r}{M_Z},
\label{eq:susythr}
\end{equation}
where the index $r$ runs over all superpartner and heavy Higgs 
particles, and $b_a^r$ represent the contributions of particle $r$ 
to the one-loop beta function coefficients \cite{Langacker:1992rq}.
Unless the supersymmetric spectrum is highly unusual, this correction
is not sufficient to bring agreement with data, and hence 
$\delta\alpha_s|_{\rm u} \neq 0$ is required.  However, the threshold 
correction at the unified scale, $\delta\alpha_s|_{\rm u}$, in general 
depends on parameters of the unified theory, and cannot be numerically 
predicted.\footnote{
The minimal supersymmetric $SU(5)$ theory is excluded because proton 
stability bounds requires $\delta\alpha_s|_{\rm u}$ to have the 
wrong sign \cite{Goto:1998qg}.}

In KK grand unified theories, the situation is quite different 
\cite{Hall:2001xb}. If the higher dimensional theory is valid over 
a sizable energy scale, up to the scale $M_s$ of strong coupling, then 
the leading unified scale corrections, coming from the KK towers, 
can be reliable computed. In the 5D $SU(5)$ theory the result for 
$\delta\alpha_s|_{\rm u}$ is given in Eq.~(\ref{eq:as-formula}). 
Together with the finite part, calculated using dimensional 
regularization \cite{Contino:2001si}, we find
\begin{equation}
  \delta\alpha_s|_{\rm u}^{\rm KK} 
    \simeq -0.0019 \, \ln\frac{M_s}{M'_c} - 0.0018 
    \simeq -0.012 \pm 0.003.
\label{eq:delta-as}
\end{equation}
Here we have used $M_s/M'_c \simeq 16\pi^3/g^2 C \simeq 200$ for our 
theory, which gives a corresponding error of $\pm 0.003$ arising from 
unknown physics above $M_s$, described in the effective theory by 
brane-localized operators at $M_s$. This contribution from KK towers 
precisely explains the difference between the experimental value, 
Eq.~(\ref{eq:as-exp}), and the central prediction without threshold 
corrections, Eq.~(\ref{eq:alphasgen}).

Breaking supersymmetry via the small boundary condition parameter 
$\alpha$ allows us to go further, since we know the superpartner 
spectrum in terms of $\tilde{m} = \alpha /R$. Substituting the 
predicted masses of Table~\ref{table:soft} into the $m_r$'s of 
Eq.~(\ref{eq:susythr}) gives
\begin{equation}
  \delta\alpha_s|_{\rm susy}^{\rm KK} 
    \simeq 0.0040 - 0.0030 \, \ln\frac{\tilde{m}}{M_Z}.
\label{eq:assusykk}
\end{equation}
Here, we have approximated the masses of the heavy Higgs bosons and 
the Higgsinos by $(2|\mu|^2 + m_{h_u}^2 + m_{h_d}^2)^{1/2}$ and 
$|\mu|$, respectively. Hence, in the wide parameter region 
$\tilde{m} = 200{\rm -}800~{\rm GeV}$, the prediction
from gauge coupling unification in our theory is
\begin{equation}
  \alpha_s^{\rm KK} = (0.116{\rm \sim}0.120) \pm 0.003.
\label{eq:askk}
\end{equation}
This is in striking agreement with data, Eq.~(\ref{eq:as-exp}).
Since threshold corrections from the weak and compactification scales
have been included, as well as the logarithmic contributions from the
KK towers, the only significant uncertainty, $\pm 0.003$, comes from 
unknown physics at and above the scale $M_s$ of strong coupling. 
While the general result for 5D KK grand unified theories is highly 
successful, in our theory we are also able to exclude the possibility 
of a large correction from the supersymmetric threshold.  

\subsection{Yukawa coupling unification}
\label{subsec:ycu}

Grand unified theories unify quarks with leptons and therefore have 
the possibility of relating quark and lepton masses. In theories where 
a single Yukawa coupling generates masses for both the $b$ quark and 
$\tau$ lepton, the mass ratio $m_b / m_\tau$ is predicted. Inputting 
the precisely known experimental value of the $\tau$ mass leads to 
a prediction for the $b$ mass
\begin{equation}
  m_b(M_Z) 
    = m_b^0 \left( 1 + \left.\frac{\delta m_b}{m_b}\right|_{\rm susy} 
      + \left.\frac{\delta m_b}{m_b}\right|_{\rm u} \right),
\label{eq:bmass}
\end{equation}
which is similar in form to the prediction for $\alpha_s(M_Z)$
of Eq.~(\ref{eq:alphasgen}) from gauge coupling unification, having 
threshold corrections from both the supersymmetry breaking and 
unified scales. An important difference is that the central prediction, 
$m_b^0$, depends on $\tan\beta$. However, in the region mainly of 
interest to us, $3 \lsim \tan\beta \lsim 20$, the dependence is weak: 
$m_b^0 = 3.6 \pm 0.1~{\rm GeV}$, where the uncertainty depends on
both $\tan \beta$ and $\alpha_s$. This should be compared with the
value of the running $b$ quark mass extracted from data
\begin{equation}
  m_b^{\rm exp} = 3.0 \pm 0.3~{\rm GeV}.
\label{eq:bmassexp}
\end{equation}
Without quark-lepton unification, there is no reason to expect the 
$b$ and $\tau$ masses to be close, hence it is significant that the 
unified Yukawa coupling leads to a broadly successful prediction. 
Nevertheless, the prediction is too large by $(10{\rm -}30)\%$. 
However, because the supersymmetric threshold corrections are typically 
large \cite{Hall:1993gn}, one usually views this as an inherent 
limitation on the sensitivity of this probe of the unified theory.
Finite one-loop diagrams with a virtual gluino and chargino give  
contributions to the bottom quark mass proportional to $\tan\beta$
\begin{equation}
  \left. \frac{\delta m_b}{m_b} \right|_{\rm susy}
    \simeq \frac{\mu \tan\beta}{4\pi} 
    \left[ \frac{8 \alpha_s m_{\tilde{g}}}{3} 
      I(m_{\tilde{g}}^2, m_{\tilde{b}_1}^2, m_{\tilde{b}_2}^2) 
      + \frac{y_t A_t}{4\pi} 
      I(|\mu|^2, m_{\tilde{t}_1}^2, m_{\tilde{t}_2}^2) \right],
\label{eq:delta-mb-Mz}
\end{equation}
where $m_{\tilde{g}}^2$ is the gluino mass, and $m_{\tilde{b}_{1,2}}$
($m_{\tilde{t}_{1,2}}$) denote the masses for the two bottom (top) 
squarks.  The function $I$ is defined by $I(a,b,c) \equiv 
-\{ ab\ln(a/b)+bc\ln(b/c)+ca\ln(c/a) \} / \{ (a-b)(b-c)(c-a) \}$. 
Hence, in most unified theories the significance of the successful 
$b$ mass prediction is limited by rather poor precision.\footnote{
The situation is not improved at large $\tan\beta$. Although $m_b^0$ 
can now approach the experimental value, the contributions to the 
supersymmetric threshold corrections are proportional to $\tan\beta$ 
and hence are large, unless cancellations occur in 
Eq.~(\ref{eq:delta-mb-Mz}).} 
In addition, there are threshold corrections from the unified scale, 
which typically cannot be calculated numerically.

The situation in our 5D $SU(5)$ KK grand unified theory is quite 
different. As in the case of gauge coupling unification, the unified 
scale corrections can be computed, and the supersymmetric corrections 
depend on only the single parameter $\alpha$.  For $m_b$ we are able 
to obtain a prediction having a smaller error bar than the current data.

\begin{figure}
\begin{center}
\begin{picture}(450,150)(-125,-20)
%
%
  \Text(-25,0)[t]{a) gauge couplings}
  \LongArrow(-120,120)(70,120)
  \Text(70,128)[b]{$\ln\mu$}
  \ArrowLine(-120,120)(-30,120)
  \Text(-110,117)[rt]{$\eta_1$}
  \DashLine(10,120)(-120,68){2}
  \ArrowLine(-120,68)(-30,104)
  \Line(-30,104)(40,120)
  \Text(-110,78)[rb]{$\eta_2$}
  \DashLine(10,120)(-120,3){2}
  \Line(40,120)(-30,93)
  \ArrowLine(-30,93)(-120,12)
  \Text(-110,26)[rb]{$\eta_3$}
  \Line(-30,116)(-30,124) \Text(-30,130)[b]{$M'_c$}
  \DashLine(-30,120)(-30,60){3}
  \DashLine(10,116)(10,124){2} \Text(10,130)[b]{$M_u$}
  \Line(40,116)(40,124) \Text(40,130)[b]{$M_s$}
%
%
  \Text(215,0)[t]{b) Yukawa couplings}
  \LongArrow(120,40)(310,40)
  \Text(310,35)[t]{$\ln\mu$}
  \DashCArc(266,230)(190,221,265){2}
  \ArrowLine(120,40)(210,40)
  \Text(125,48)[b]{$\zeta_\tau$}
  \ArrowArcn(257,214)(180,-105,221)
  \Text(125,113)[b]{$\zeta_b$}
  \Line(210,36)(210,44) \Text(210,30)[t]{$M'_c$}
  \DashLine(250,36)(250,44){2} \Text(250,30)[t]{$M_u$}
  \Line(280,36)(280,44) \Text(280,30)[t]{$M_s$}
\end{picture}
\caption{a) The running of the difference of the three gauge 
 couplings, $\eta_i \equiv \alpha_i^{-1} - \alpha_1^{-1}$.
 Solid and dashed lines represent the runnings in KK grand unification 
 and conventional 4D grand unification, respectively.  
 In KK grand unification, $\eta_i$'s run logarithmically both below 
 and above $M'_c$, but with different beta function coefficients, 
 so that the prediction for the QCD coupling differs from that in the 
 case of a single scale unification.  
 b) The running of the Yukawa couplings, $\zeta_f \equiv 
 y_f/y_\tau - 1$ in KK grand unification (solid) and 4D grand 
 unification (dashed).  In KK grand unification there is no 
 $SU(5)$-violating running above $M'_c$, so that the predicted value 
 of the bottom quark mass is smaller than that of the single scale 
 unification case.}
\label{fig:running}
\end{center}
\end{figure}
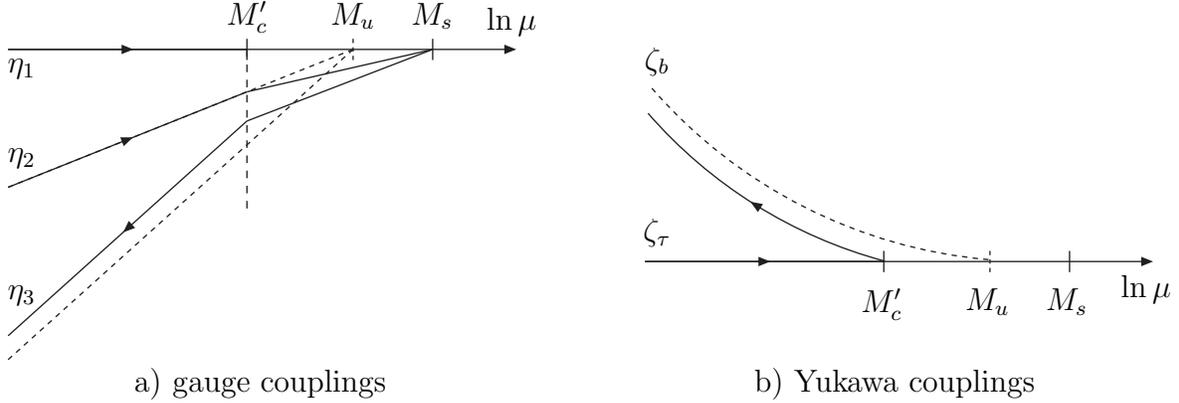

First, we consider the correction from the unified scale.
The largest effect in our theory arises because the bottom 
and tau Yukawa couplings are unified at the compactification scale 
$M'_c$, which is smaller than the conventional unification scale 
$M_u \simeq 2 \times 10^{16}~{\rm GeV}$. This is different from the
situation in gauge coupling unification, where the couplings unify at
$M_s$ rather than $M'_c$. It is interesting to understand the different
behavior of gauge and Yukawa unification. The renormalization group 
running below and above $M'_c$ can have completely different origins 
in the higher dimensional picture.  Below $M'_c$ the running can be 
caused by the generation of operators which are non-local along the 
extra dimension, while above $M'_c$ all the running must come from 
local operators in the extra dimension.  In the case of the Yukawa 
couplings, the bottom and tau Yukawa couplings evolve differently below 
the compactification scale.  This is obvious in the 4D picture 
because the effective theory below $M'_c$ is just the usual MSSM, 
but in the 5D picture it is not so obvious, because both $T_3$ and 
$F_3$ are localized on the $y=0$ brane where the complete $SU(5)$
symmetry is operative.  Nevertheless there is a simple understanding: 
below $M'_c$, radiative corrections generate kinetic operators for 
$T_3$ and $F_3$ that are non-local along the extra dimension (effective 
kinetic operators involving Wilson lines or Wilson loops).  Since these 
operators are non-local (i.e. go around the circle), they feel $SU(5)$ 
violating effects from the boundary conditions, and hence are $SU(5)$ 
violating.  Clearly, the effects from these operators exponentially 
shut off above $M'_c$ because of their non-local nature.  Thus whether 
there is $SU(5)$ violating running at energies above $M'_c$ or not is 
determined by whether we can write down an $SU(5)$ violating local 
operator or not.  However, in our theory both $T_3$ and $F_3$ are 
localized on $SU(5)$-invariant $y=0$ brane, so their kinetic and Yukawa 
interactions cannot violate $SU(5)$ --- above $M'_c$ the bottom and 
tau Yukawa couplings must run together and hence are unified.
This is in contrast with the case of gauge coupling unification, where 
the gauge fields propagate in the bulk and can have $SU(5)$ violating 
kinetic energy operators located on the $y=\pi R$ brane.  As a result, 
gauge couplings have $SU(5)$ violating logarithmic evolution above 
$M'_c$, while the third generation Yukawa couplings do not, as 
illustrated in Fig.~\ref{fig:running}.

The correction from the unified scale, $\delta m_b/m_b|_{\rm u}$, 
is calculated as follows.  At leading order, the bottom and tau Yukawa 
couplings are split because they receive different contributions 
from QCD, hypercharge and top Yukawa radiative corrections below their 
unification scale.  Since the unification scale in our theory, $M_c'$, 
differs from that in conventional 4D supersymmetric unified theories, 
$M_u$, we obtain the correction
\begin{equation}
  \left. \frac{\delta m_b}{m_b} \right|_{\rm u}^{\rm KK}
    \simeq - \frac{4g^2 - y_t^2}{16\pi^2} \ln\frac{M_u}{M'_c},
\label{eq:delta-mb-Mc}
\end{equation}
where $g$ and $y_t$ represent the values of the 4D gauge coupling and 
the top Yukawa coupling at the compactification scale, $g \simeq 0.7, 
y_t \simeq 0.6$.  Note that the gauge contribution dominates over the 
top contribution so that the net effect is to reduce the prediction for 
$m_b(M_Z)$ and therefore reduce the discrepancy with data.  This result 
is independent of how supersymmetry is broken and has general 
applicability to all KK grand unified theories having $b/\tau$ 
unification. Using Eq.~(\ref{eq:mc-formula}), we obtain 
$\delta m_b/m_b|_{\rm u}^{\rm KK} \simeq 
-(5(4 g^2 - y_t^2)/112\pi^2)\ln(M_s/M'_c)$, so that for our theory, 
$M_s/M'_c \approx 200$, we find $\delta m_b/m_b|_{\rm u}^{\rm KK} 
\approx -4\%$.

Substituting predictions for $\mu$ and the superpartner masses from 
Table~\ref{table:soft} into the expression Eq.~(\ref{eq:delta-mb-Mz}) 
for the supersymmetric threshold correction, we find
\begin{equation}
  \left. \frac{\delta m_b}{m_b} \right|_{\rm susy}^{\rm KK}
    \simeq 0.006 \times {\rm sgn}(\mu) \tan\beta.
\label{eq:delta-mb-Mz-num}
\end{equation}
Strictly speaking, the mass eigenvalues for the bottom and top squarks 
depend on the ratio $v/\tilde{m}$; however, we have ignored this and 
find Eq.~(\ref{eq:delta-mb-Mz-num}) gives a good approximate numerical 
result over a wide range of the parameter space.  Choosing 
${\rm sgn}(\mu) < 0$, the prediction for the $b$ quark mass in our 
theory is
\begin{equation}
  m_b(M_Z) = 3.3 - 0.02 \, (\tan\beta - 10) \pm 0.1~{\rm GeV}.
\label{eq:mbpred}
\end{equation}
While a more precise form for the prediction in terms of $\tan\beta$
and $\alpha_s$ is possible, it is not warranted by the present error
from extracting $m_b$ from data. The simple prediction of
Eq.~(\ref{eq:mbpred}) is nevertheless quite important. Our KK theory 
does bring the prediction for $m_b/m_\tau$ into agreement with data
without the need to invoke unknown threshold correction of at least 
$10\%$. If the corrections from KK modes had gone in the wrong direction, 
our theory would have been very clearly excluded. Instead we predict 
that $m_b$ lies at the upper edge of the presently allowed experimental 
region.\footnote{
In the case where small neutrino masses are generated through the 
see-saw mechanism, the presence of Dirac neutrino Yukawa couplings 
at high energies could increase the prediction for $m_b$ by a small 
amount but does not change our conclusions.  The $U(1)_X$ gauge 
interaction does not renormalize $m_b/m_\tau$ and hence does not 
affect our prediction.}

\subsection{Proton decay}
\label{subsec:pdecay}

In KK grand unified theories proton decay from operators of 
dimension four and five is forbidden \cite{Hall:2001pg}, so that the 
only possible source of significant proton decay is from operators of 
dimension six.  Since brane-localized dimension six operators, 
with coefficients given by inverse powers of $M_s$, give effects 
which are too small to observe, the only possible relevant source 
is from exchange of the superheavy gauge bosons, such as the $X$ 
boson of $SU(5)$.  In 4D supersymmetric unified theories, the 
$X$ gauge boson exchange contribution is negligible since the $X$ 
gauge boson mass is $M_X \simeq M_u \simeq 2 \times 10^{16}~{\rm GeV}$. 
However, in KK unified theories gauge coupling unification results only 
if the volume of the bulk is large, and this leads to lower values for 
$M_X = M_c/2$, making the gauge boson exchange contribution much more 
important \cite{Hall:2001pg}. A precise prediction for gauge coupling 
unification results when the volume of the bulk is as large as possible, 
so that the gauge interactions at $M_s$ are strong, and this leads 
to low and calculable values for $M_X$, opening the possibility 
for interesting predictions for the proton decay rate 
\cite{Nomura:2001tn, Hall:2001xb}.

In fact, while gauge boson mediated proton decay is of great 
interest in KK grand unified theories, at present significant 
uncertainties remain.  Exchange of $X$ gauge bosons results in 
baryon number violating four fermion operators of the form 
$[T_i^\dagger T_i T_j^\dagger T_j]_{\theta^2 \bar{\theta}^2}$ and 
$[T_i^\dagger T_i F_j^\dagger F_j]_{\theta^2 \bar{\theta}^2}$.  The 
baryon number violating interactions amongst zero modes are generated 
only when $T$ and $F$ are brane fields, since for bulk fields the $X$ 
boson couples a zero mode to a superheavy state.  The coefficients 
for these operators are given by $\sum_n g_X^2/((n+1/2)M_c)^2 
= g^2/M_c^{\prime\, 2}$, where $g_X$ and $g$ are the gauge couplings 
for the $X$ and standard model gauge bosons, respectively, and 
$g_X = \sqrt{2} g$ due to the wavefunction profiles of the gauge fields 
in the extra dimension.  Since the corresponding coefficients are given 
by $g_X^2/M_X^2 = g^2/M_X^2$ in 4D grand unified theories, $M'_c$ in 5D 
theories corresponds to $M_X$ in 4D theories as far as dimension 
six proton decay is concerned.  For $M'_c \approx 10^{15}~{\rm GeV}$, 
present limits on proton decay require that $T_1$ be a bulk field. 
The bulk gauge interactions then generate proton decay 
through quark mass mixing and/or brane-localized interactions 
\cite{Nomura:2001tn, Hebecker:2002rc}.  This introduces uncertainties 
for the proton decay rate, since the relevant mixing matrix is not the 
CKM matrix of weak interactions and the size of brane interactions 
are not determined by gauge coupling unification. Nevertheless, it is 
still interesting to pursue the natural expectation for the rate for 
various proton decay modes in these KK grand unified theories.

In our theory $F_i$ are all brane fields, but this is not sufficient to 
generate baryon number violation.  Since only the third generation is 
on the brane for $T$ fields, proton decay is dominated by the operator
$[T_3^\dagger T_3 F_1^\dagger F_1]_{\theta^2 \bar{\theta}^2}$ and hence
requires mixing via the third generation of quarks.  We then find that 
the four fermion interaction ${\cal L} \supset (g^2/M_c^{\prime\, 2})
(V^{q\dagger}_{23} V^u_{31}) q_2^\dagger u_1 l_1^\dagger d_1$ gives 
the dominant proton decay amplitude, where $V^{q,u}$ is the unitary 
rotation on $q,u$ needed to diagonalize the quark mass matrices. 
Since we must take the strange quark from $q_2$ to obtain a proton 
decay amplitude, $V^q$ is actually the rotation in the left-handed 
down quark sector and the final states for the decay contain the 
particle carrying the strangeness quantum number.  This leads to the 
decay mode $p \rightarrow K^+ \bar{\nu}$ with the lifetime estimate 
$\tau(p \rightarrow K^+ \bar{\nu}) \approx 10^{37 \pm 2}~{\rm years}$. 
Here, we have arbitrarily assigned one order of magnitude uncertainty 
to $M_c^{\prime\, 4}$ and one to the flavor mixing angles; the central 
value comes from taking $V^{q\dagger}_{23} V^u_{31} \simeq 0.0002$ and 
$M'_c = M_u (M'_c/M_s)^{5/7}$ with $M_s/M'_c \simeq 16\pi^3/g^2 C 
\simeq 200$.  The simple expectation is therefore that this flavor 
mixing contribution to proton decay is too small to observe. In general 
in KK unified theories one only expects an observable rate if $T_2$ 
is on the brane \cite{Nomura:2001tn, Hall:2001xb}, or if $M'_c$ is 
of order $10^{14}~{\rm GeV}$ rather than of order $10^{15}~{\rm GeV}$ 
\cite{Hebecker:2002rc}.

Symmetries of our theory allow the operators 
$[(\bar{g}_{ij}/M_s) T_i^\dagger T'_j]_{\theta^2 \bar{\theta}^2}$ 
on the $y=0$ brane.  After the KK mode decomposition, these operators 
give baryon number violating couplings amongst zero modes: ${\cal L} 
\supset \bar{g}_{ij} (M'_c/M_s) u_i^\dagger \gamma^\mu q_j X_\mu$.
The leading proton decay in our theory comes from combining these 
interactions with the $d_i^\dagger \gamma^\mu l_i X_\mu$ interactions 
coming from the usual kinetic terms $[F_i^\dagger F_i]_{\theta^2 
\bar{\theta}^2}$ on the brane.  The resulting decay amplitudes are 
independent of the Yukawa interactions (i.e. quark mixing matrices), 
but depend on the unknown coefficients $\bar{g}_{ij}$.  We find that 
$\bar{g}_{22}$ and $\bar{g}_{21}$ do not lead to proton decay 
amplitudes at leading order in $(M'_c/M_s)$, but $\bar{g}_{12}$ 
and $\bar{g}_{11}$ lead to $p \rightarrow K^+\bar{\nu}$ and 
$p \rightarrow e^+\pi^0, \mu^+\pi^0, e^+K^0, \mu^+K^0, \pi^+\bar{\nu}, 
K^+\bar{\nu}$ decay modes, respectively, both at order 
$(M'_c/M_s)$ in the amplitudes.  A crucial question is whether 
the origin of the $d/s, u/c$ ratios and the Cabibbo angle leads to 
further suppression of $\bar{g}_{11,12}$ relative to $\bar{g}_{22}$.  
In any of the examples given in sub-section \ref{subsec:masses}, 
there is no such suppression for $\bar{g}_{11}$, so we expect the 
coefficient $\bar{g}_{11}$ to be of order one.  This gives the 
``effective $X$ gauge boson mass'' $M_X^{\rm eff} \simeq 
\bar{g}_{11}^{-1/2} (M'_c/M_s)^{3/14} M_u$, leading to the lifetime 
estimate $\tau(p) \approx 10^{34}~{\rm years}$ for all the decay 
modes $e^+\pi^0, \mu^+\pi^0, e^+K^0, \mu^+K^0, \pi^+\bar{\nu}$ and 
$K^+\bar{\nu}$, for $\bar{g}_{11} \simeq 1$ and $M_s/M'_c \simeq 200$.
An interesting point is that, although the total proton decay rate has 
uncertainties coming from $\bar{g}_{11}$, the relative decay rates for 
various decay modes are predicted essentially in terms of a single 
unknown parameter $\theta_F$.  This is because we can always choose 
a gauge eigenbasis for $F_i$ so that only $F_3$ couples to $T_3$, and 
in that basis the only large mixing angle for quarks and leptons in 
$F_i$ is that between the first two generations (see also the discussion 
around Eq.~(\ref{eq:yde})).  We define the relative mixing angles 
between the down type quarks and the lepton doublets as $\theta_F$, 
which we expect to be of order unity.  We then obtain a number of 
predictions for the relative decay rates, so that the physics of proton 
decay is very rich in our theory.  Particularly useful relations are 
\begin{equation}
  \frac{\Gamma(p \rightarrow \mu^+\pi^0)}{\Gamma(p \rightarrow e^+\pi^0)} 
\simeq 
  \frac{\Gamma(p \rightarrow e^+K^0)}{\Gamma(p \rightarrow \mu^+K^0)} 
\simeq \tan^2\theta_F.
\label{eq:pratios}
\end{equation}
This is a robust prediction because it does not depend on hadronic 
matrix elements nor whether $\bar{g}_{12}$ is sizable.  For the first 
discovery, $e^+\pi^0$ will be the most promising mode for an experimental 
search, since it has a relatively clean signature.  We stress that, 
while our analysis for proton decay depends on matter location, it is 
completely independent of supersymmetry breaking.

\subsection{Axion and axino from $U(1)_R$}
\label{subsec:r-axion}

The bulk interactions of our theory possess an $SU(2)_R \times SU(2)_H$ 
global symmetry, and we have argued that it is crucial for the $U(1)_R$ 
subgroup to be preserved also by all the brane interactions.  On the 
other hand the small boundary condition parameter $\alpha$ clearly 
breaks this $U(1)_R$ symmetry. However, this breaking will ultimately 
be spontaneous, arising from the vacuum expectation value of an 
auxiliary field in the 5D supergravity multiplet. It is therefore 
natural to explore the possibility that the global $U(1)_R$ symmetry is 
an exact symmetry of the entire theory, except for gauge anomalies, 
so that an $R$ axion arises from the spontaneous breaking of $U(1)_R$. 
An interesting possibility is that the $R$ axion only receives mass 
from the gauge interactions of the standard model.  (In fact, in 
section \ref{sec:develop} we will introduce an additional gauge 
interaction which sets the $U(1)_X$ breaking scale, but $U(1)_R$ does 
not have an anomaly for this gauge group so that the $R$ axion does not 
get a mass from this gauge interaction.)  If this is the case, the 
$R$ axion receives the dominant mass contribution from the QCD anomaly 
of $U(1)_R$ and thus plays the role of the QCD axion, solving the 
strong $CP$ problem \cite{Peccei:1977hh}. The $U(1)_R$ breaking scale 
is given (if these is no other $U(1)_R$ breaking) by the supersymmetry 
breaking vacuum expectation value of the auxiliary field: 
$F_a \approx (\tilde{m}M_{\rm Pl})^{1/2} \approx 
10^{10}{\rm -}10^{11}~{\rm GeV}$.\footnote{
If a constant term in the superpotential canceling the cosmological 
constant is generated by a vacuum expectation value as 
$W = \vev{\Phi}^3$, then the scale of $U(1)_R$ breaking is 
given by $F_a \approx (\tilde{m}M_{\rm Pl}^2)^{1/3} \approx 
10^{13}~{\rm GeV}$.}

We here discuss possible phenomenological consequences of this exact 
$U(1)_R$ scenario.  First, the mass of the (fermionic) superpartner of 
the $R$ axion, the $R$ axino, strongly depends on the sector which 
breaks $U(1)_R$ spontaneously.  In particular, the $R$ axino could be 
much lighter than all the superpartners of the standard model particles. 
In this case, the lightest standard model superpartner can decay into 
the $R$ axino, so that there is no problem of charged dark matter even 
if the stau is the lightest of the MSSM superparticles.  The lifetimes 
for stau and neutralino decays are estimated roughly to be 
$10^{3}(F_a/10^{11}~{\rm GeV})^2 (10^2~{\rm GeV}/
m_{\tilde{\tau}})^3~{\rm sec}$ and $10^{-3}(F_a/10^{11}~{\rm GeV})^2 
(10^2~{\rm GeV}/m_{\tilde{\chi}})^3~{\rm sec}$, respectively.  The dark 
matter may be provided by the $R$ axion or by the $R$ axino itself.

\section{Supersymmetric Flavor Violation}
\label{sec:flavor}

In this section we propose lepton flavor violation as a powerful and 
generic signal for KK grand unification providing the scale of mediation 
for supersymmetry breaking is at or above the compactification scale.
We first consider the general consequences for flavor symmetry imposed 
by matter locality and experimental limits on lepton flavor violation.
We then apply this to our model and derive rates for various signals 
in the lepton sector.  We argue that hadronic signals are likely to be 
harder to detect.

\subsection{Flavor symmetry in Kaluza-Klein grand unification}
\label{subsec:flavor-symm}

In supersymmetric theories, flavor violation provides an important probe 
of the high energy theory via the form of the soft supersymmetry breaking
interactions \cite{Hall:1985dx}. Furthermore, the form of the soft 
operators reflects the underlying flavor symmetry of the theory. For 
example, in conventional 4D $SU(5)$ grand unified theories, the gauge 
interactions possess a $U(3)_T \times U(3)_F$ global flavor symmetry. 
If soft supersymmetry breaking operators are local up to the unification 
scale, in the flavor symmetry limit the squark and slepton mass matrices 
have the form $m_{\tilde{T}}^2 (\tilde{q} \tilde{q}^\dagger + 
\tilde{u}^\dagger \tilde{u} + \tilde{e} \tilde{e}^\dagger) + 
m_{\tilde{F}}^2 (\tilde{d}^\dagger \tilde{d} + \tilde{l}^\dagger 
\tilde{l})$. Here $\tilde{q},\tilde{e}$ ($\tilde{u},\tilde{d},\tilde{l}$) 
are three dimensional row (column) vectors in flavor space. Flavor 
violating signals only arise from the breaking of flavor symmetry due to
non-gauge interactions, such as the top quark Yukawa coupling.

On the other hand, in KK grand unified theories the flavor symmetry 
of the gauge interactions depends on the location of the matter. In 5D 
theories with a large radius of the extra dimension, we have seen that 
$T_1$ and $T_3$ have differing locations, so that $U(3)_T$ is certainly 
broken. If soft supersymmetry breaking operators are generated at or 
above the compactification scale this implies a large mass splitting 
between the scalars in $T_1$ and those in $T_3$. However, constraints 
from flavor violating processes place powerful limits on the degree to 
which $U(3)_T \times U(3)_F$ can be ``broken'' by locality. For example, 
$T_1$ and $T_2$ must have a common location, as must $F_{1,2,3}$, to 
avoid too large lepton flavor violation, as will be seen later. Hence 
in KK grand unified theories with a ``large'' extra dimension we find 
that the most general form for the flavor symmetry is 
$(U(2) \times U(1))_U \times (U(2) \times U(1))_E \times 
(U(2) \times U(1))_Q \times U(3)_D \times U(3)_L$, leading to 
the soft scalar masses
\begin{equation}
  -{\cal L} = m_{\tilde{q}}^2 \tilde{q} P_q \tilde{q}^\dagger
    + m_{\tilde{u}}^2 \tilde{u}^\dagger P_u \tilde{u} 
    + m_{\tilde{d}}^2 \tilde{d}^\dagger \tilde{d} 
    + m_{\tilde{l}}^2 \tilde{l}^\dagger \tilde{l} 
    + m_{\tilde{e}}^2 \tilde{e} P_e \tilde{e}^\dagger,
\label{eq:sq-sl-m2}
\end{equation}
where $P_{q,u,e}$ are $3 \times 3$ matrices $P_{q,u,e} = 
{\rm diag}(1,1,c_{q,u,e})$ containing arbitrary parameters 
$c_{q,u,e}$.\footnote{If $c_e$ happens to be close to unity, $F_{1,2}$ and 
$F_3$ could have different locations and the flavor symmetry could be 
$(U(2) \times U(1))^5$.}
Therefore, unless $c_q = c_u = c_e = 1$, the soft supersymmetry 
breaking masses are not flavor universal even in the ``flavor 
symmetric'' limit. Here, by allowing $m_{\tilde{q}}^2, m_{\tilde{u}}^2, 
m_{\tilde{d}}^2, m_{\tilde{l}}^2, m_{\tilde{e}}^2$ (and $c_q, c_u, c_e$) 
to be independent, we have allowed for the case that supersymmetry 
breaking directly feels the breaking of gauge symmetry. This can 
occur easily in KK unified theories since there are locations in the 
bulk which have explicit breaking of the unified symmetry.
The three standard model gaugino masses must also be treated as 
independent parameters in general in KK grand unified theories.
In the case that supersymmetry breaking respects the $SU(5)$ symmetry, 
the flavor symmetry is reduced to $U(1)_{T_3} \times U(2)_{(U,E)_{1,2}} 
\times U(2)_{Q_{1,2}} \times U(3)_D \times U(3)_L$ and 
Eq.~(\ref{eq:sq-sl-m2}) holds with $m_{\tilde{u}}^2 = 
m_{\tilde{e}}^2$, $c_u = c_e$ and $m_{\tilde{q}}^2 c_q = 
m_{\tilde{u}}^2 c_u$. The gaugino masses respect the unified mass 
relations. This is the case in our boundary condition 
supersymmetry breaking. Much of the remaining parameter freedom then 
arises because bulk matter is not fully unified; for instance, 
if $F_i$ are brane fields then we also have $m_{\tilde{d}}^2 = 
m_{\tilde{l}}^2$. However, these additional restrictions on parameters 
do not eliminate flavor non-universalities, since we still have 
parameters $c_q$ and $c_e$, breaking flavor universality. Hence, 
from flavor symmetry grounds alone we see that KK grand unified 
theories with supersymmetry breaking mediated at or above the 
compactification scale generically have large lepton flavor violation 
originating from the split in location between $\tilde{e}_3$ and 
$\tilde{e}_{1,2}$. Below we calculate the rates in our theory --- 
but the signal is generic to a wide class of KK unified theories with 
high mediation scales.  In fact, the signal does not require grand 
unification: a separation of lepton locations is sufficient. In the 
unified case this separation is required, but even in non-unified 
theories such a separation may well contribute to fermion mass 
hierarchies.  If supersymmetry breaking is mediated at scales well 
beneath the compactification scale, $U(3)^5$ flavor symmetry can 
emerge as an accidental low energy symmetry despite its short 
distance breaking by matter locality, removing the signal.

The flavor symmetry in our theory is $U(1)_{T_3} \times 
U(2)_{T_{1,2}} \times U(2)_{T'_{1,2}} \times U(3)_F$, due to the 
unique choice for the location of matter, shown in Fig.~\ref{fig:theory}. 
This flavor symmetry determines the overall flavor structure of the 
theory, especially that of supersymmetry breaking parameters.
To understand the consequences of this flavor symmetry, we first 
consider the limit where all the Yukawa couplings are set equal to zero.
Boundary condition breaking of supersymmetry by the parameter $\alpha$ 
generates non-holomorphic scalar masses for squarks, sleptons and 
Higgs bosons.  Since the boundary condition breaking does not introduce 
any additional flavor symmetry breaking, these supersymmetry breaking 
operators must respect the flavor symmetry of the gauge sector, giving
soft masses of the form Eq.~(\ref{eq:sq-sl-m2}) with $m_{\tilde{u}}^2 = 
m_{\tilde{e}}^2$, $c_u = c_e$, $m_{\tilde{q}}^2 c_q = 
m_{\tilde{u}}^2 c_u$ and $m_{\tilde{d}}^2 = m_{\tilde{l}}^2$.
Thus, if $c_q$ or $c_e \neq 1$, the soft supersymmetry breaking masses 
are not flavor universal even in the limit of vanishing Yukawa 
couplings. In fact, in the present theory, $m_{\tilde{q}}^2 =
m_{\tilde{u}}^2 = m_{\tilde{e}}^2 = \tilde{m}^2$, $m_{\tilde{d}}^2 = 
m_{\tilde{l}}^2 = 0$ and $c_q = c_u = c_e = 0$ at the compactification 
scale, and thus there are flavor non-universalities already at this 
level. This situation is quite different from that of conventional 
4D unified theories where all the flavor non-universalities come 
from renormalization effects through the Yukawa couplings.

We next turn to the Yukawa couplings. The flavor symmetries of the 
5D gauge interactions are broken by the brane-localized Yukawa 
couplings of Eq.~(\ref{eq:yukawa}). Flavor hierarchies for quark and 
lepton masses and mixings arise from geometrical volume suppressions, 
as shown in Eq.~(\ref{eq:fermion-yukawa}), explaining much of the gross 
structure of these hierarchies. In the low energy theory below the
compactification scale, this flavor symmetry breaking is represented
by the three superpotential Yukawa interactions
\begin{equation}
  W = Q {\bf y_U} U H_u - Q {\bf y_D} D H_d - E {\bf y_E} L H_d,
\label{eq:4dyukawas}
\end{equation}
where ${\bf y_U}$ (${\bf y_{D,E}}$) is a $3 \times 3$ matrix and has 
roughly the form of $y_T$ ($y_F$) shown in Eq.~(\ref{eq:fermion-yukawa}); 
$Q,E$ ($U,D,L$) are three dimensional row (column) vectors in the flavor 
space. Since the soft scalar masses for $F_i$ (i.e. $D_i$ and $L_i$) are 
flavor universal, we can choose a basis for $F_i$ so that only $F_3$ 
couples to $T_3$ by rotating full supermultiplets $F_i$, giving 
\begin{equation}
  {\bf y_{D,E}} \approx \pmatrix{
    \epsilon &  \epsilon &  \epsilon      \cr
    \epsilon &  \epsilon &  \epsilon      \cr
    0 & 0 & 1      \cr  }.
\label{eq:yde}
\end{equation}
This shows that there is no large mixing effect in the charged lepton 
sector, in spite of the presence of apparent large mixing angles in 
Eq.~(\ref{eq:fermion-yukawa}).  This remains true no matter which 
mechanism is used to provide a hierarchy between the first two 
generations.  The large mixing angles appear in physical processes 
only when we consider the masses for neutrinos; in particular, they 
appear as large mixing angles in neutrino oscillation phenomena.

To summarize, the soft supersymmetry breaking operators in our theory 
involve a flavor non-universality matrix $P_T$ associated with the 
states $Q,U,E$ of $T$:
\begin{equation}
  P_T = \pmatrix{
     1 & 0 & 0  \cr
     0 & 1 & 0  \cr
     0 & 0 & 0  \cr }.
\label{eq:P}
\end{equation}
The flavor universality is also broken by the Yukawa couplings, which 
we can view as flavor symmetry breaking spurions.  Applying the general 
result Eq.~(\ref{eq:soft}) for the soft operators to our theory, we 
find the following soft terms for squarks and sleptons:\footnote{
The most general form of the trilinear scalar interactions consistent 
with the flavor symmetry of the 5D gauge interactions is $-{\cal L} 
= \tilde{q} (A_u {\bf y_U} + A'_u P_q {\bf y_U} + A''_u {\bf y_U} P_u) 
\tilde{u} h_u - \tilde{q} (A_d {\bf y_D} + A'_d P_q {\bf y_D}) 
\tilde{d} h_d - \tilde{e} (A_e {\bf y_E} + A'_e P_e {\bf y_E}) 
\tilde{l} h_d$.}
\begin{eqnarray}
  {\cal L}_{\rm soft} 
  &=& \tilde{m} \left( 
     \tilde{q} ({\bf y_U} + P_T {\bf y_U} + {\bf y_U} P_T) \tilde{u} h_u
     - \tilde{q} ({\bf y_D} + P_T {\bf y_D}) \tilde{d} h_d
     - \tilde{e} ({\bf y_E} + P_T {\bf y_E}) \tilde{l} h_d \right) 
\nonumber \\
  && - \tilde{m}^2 \left( \tilde{q} P_T \tilde{q}^\dagger
     + \tilde{u}^\dagger P_T \tilde{u} 
     + \tilde{e} P_T \tilde{e}^\dagger \right).
\label{eq:soft2}
\end{eqnarray}
This tree-level result corresponds to the initial condition at 
the compactification scale for the renormalization group scaling 
of the soft parameters of the MSSM.

\subsection{Lepton flavor violation}
\label{subsec:lepton-flavor}

\subsubsection{A single lepton flavor mixing matrix}
\label{subsubsec:we}

To study the experimental consequences of lepton flavor violation
in our theory, we must first scale the operators $[EL H_d]_{\theta^2}$, 
$\tilde{e}\tilde{l} h_d, \tilde{l}^\dagger \tilde{l}$ and 
$\tilde{e} \tilde{e}^\dagger$ to the weak scale.  In the case that 
$\tan\beta$ is not too large, so that radiative effects from 
$y_{b,\tau}$ can be ignored, the only important scalings are due to 
the electroweak gauge interactions. Thus the overall form of the 
Yukawa matrix ${\bf y_E}$ is unchanged and has the form Eq.~(\ref{eq:yde}) 
also at the weak scale. There is a radiative correction $\delta A$ to 
the trilinear scalar operator proportional to the gaugino mass, and 
there are gauge radiative corrections to the soft scalar mass 
parameters, giving
\begin{eqnarray}
  {\cal L}_{\rm soft}^{\rm lept} 
  &\simeq& -\tilde{e} \left[ \pmatrix{
                2\tilde{m} + \delta A & 0 & 0  \cr
                0 & 2\tilde{m} + \delta A & 0  \cr
                0 & 0 & \tilde{m} + \delta A   \cr } 
      {\bf y_E} \right] \tilde{l} h_d + {\rm h.c.}
\nonumber\\
  && - \tilde{e} \pmatrix{
                m_{\tilde{e}}^2 & 0 & 0    \cr
                0 & m_{\tilde{e}}^2 & 0    \cr
                0 & 0 & m_{\tilde{\tau}}^2 \cr } \tilde{e}^\dagger
  - \tilde{l}^\dagger \pmatrix{
                m_{\tilde{l}}^2 & 0 & 0  \cr
                0 & m_{\tilde{l}}^2 & 0  \cr
                0 & 0 & m_{\tilde{l}}^2  \cr } \tilde{l},
\label{eq:softlept}
\end{eqnarray}
where, using Table~\ref{table:soft}, $m_{\tilde{e}}^2 = 
m_{\tilde{e}_B}^2 \simeq (1.1\,\tilde{m})^2$, $m_{\tilde{\tau}}^2 = 
m_{\tilde{e}_3}^2 \simeq (0.33\,\tilde{m})^2$, 
$m_{\tilde{l}}^2 = m_{\tilde{l}_b}^2 \simeq m_{\tilde{l}_3}^2 
\simeq (0.70\,\tilde{m})^2$ and $\delta A = 0.60\,\tilde{m}$.

By rotating to a mass eigenstate basis for charged leptons, while 
maintaining diagonal scalar mass-squared matrices, we can go to a basis 
where the lepton flavor violation appears only via a single new 
mixing matrix $W^e$ in the following lepton-slepton-gaugino, 
slepton-lepton-Higgsino and slepton-slepton-Higgs interactions:
\begin{eqnarray}
  {\cal L}^{\rm LFV} &=& - \left( \sqrt{2}g' e W^{e\dagger} 
      \tilde{e}^\dagger \tilde{b} + {\rm h.c.} \right)
    + \left( \tilde{e} W^e {\bf \hat{y}_E} l \tilde{h}_d + {\rm h.c.} \right)
\nonumber\\
 && - \left( \tilde{e} \left( (\tilde{m}+\delta A)I + \tilde{m}P_T \right)
      W^e {\bf \hat{y}_E} \tilde{l} h_d 
    - \mu\, \tilde{e} W^e {\bf \hat{y}_E} h_u^\dagger \tilde{l} 
    + {\rm h.c.} \right),
\label{eq:lfvwe}
\end{eqnarray}
where ${\bf \hat{y}_E}$ is the real and diagonal lepton Yukawa matrix 
after the rotation, and $I$ is the unit $3 \times 3$ matrix; 
$\tilde{b}$ represents the $U(1)_Y$ gaugino.  Note that, because the 
mass-squared matrix for $\tilde{l}$ is proportional to the unit matrix, 
the lepton flavor violation is associated with $\tilde{e}$ rather than 
with $\tilde{l}$.\footnote{
This is similar to lepton flavor violation in 4D supersymmetric $SU(5)$ 
theories with $\tilde{e}$ non-degeneracy arising from radiative 
corrections involving the large coupling $y_T$ for the third 
generation \cite{Barbieri:1994pv}.} 
The degeneracy of $\tilde{e}_{1,2}$ allows $W^e$ to depend on only 
two physical Euler angles and two phases as $W^e = R^e_{23} R^e_{12} D$, 
where $R^e_{ij}$ is a matrix rotating the $ij$ plane in flavor space, 
and $D$ is a diagonal phase matrix with two independent phases. 
A simultaneous phase rotation for $l$, $\tilde{l}$ and $e^\dagger$ 
(i.e. $l \rightarrow D^\dagger l$, $\tilde{l} \rightarrow D^\dagger 
\tilde{l}$ and $e^\dagger \rightarrow D^\dagger e^\dagger$) can 
further remove the phase matrix $D$ from $W^e$ without affecting 
other interactions, so that we finally obtain 
\begin{equation}
  W^e = R^e_{23} R^e_{12} 
    = \pmatrix{
          c^e_{12}           & -s^e_{12}          & 0          \cr
          s^e_{12}\,c^e_{23} & c^e_{12}\,c^e_{23} & -s^e_{23}  \cr
          s^e_{12}\,s^e_{23} & c^e_{12}\,s^e_{23} & c^e_{23}   \cr },
\label{eq:we}
\end{equation}
where $c^e_{ij} \equiv \cos(\theta^e_{ij})$ and 
$s^e_{ij} \equiv \sin(\theta^e_{ij})$.  Therefore, we find a remarkable 
result that all the lepton flavor violating processes are completely 
described by two angles, $\theta^e_{12}$ and $\theta^e_{23}$, as far as 
the charged lepton sector is concerned.

\subsubsection{Branching ratio for $\mu \rightarrow e \gamma$}
\label{subsubsec:mu-e-gamma}

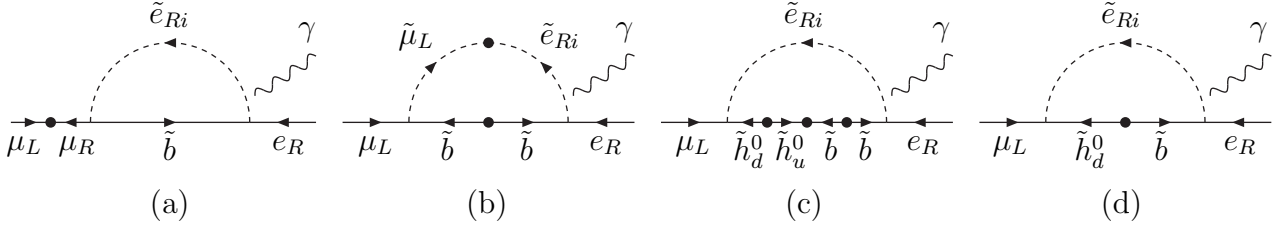
\begin{figure}
\begin{center}
\begin{picture}(480,80)(-21,10)
  \Text(40,25)[t]{(a)}
  \ArrowLine(-20,50)(-5,50) \Text(-15,45)[t]{$\mu_L$}
  \ArrowLine(10,50)(-5,50)  \Text(5,45)[t]{$\mu_R$}  \Vertex(-5,50){2}
  \ArrowLine(10,50)(70,50)  \Text(40,47)[t]{$\tilde{b}$}
  \DashArrowArc(40,50)(30,0,180){2} \Text(40,88)[b]{$\tilde{e}_{Ri}$}
  \ArrowLine(95,50)(70,50)  \Text(85,45)[t]{$e_R$}
  \Photon(72,60)(95,75){2}{3.5} \Text(92,83)[b]{$\gamma$}
  \Text(160,25)[t]{(b)}
  \ArrowLine(105,50)(130,50) \Text(118,45)[t]{$\mu_L$}
  \ArrowLine(160,50)(130,50) \Text(145,47)[t]{$\tilde{b}$}
  \ArrowLine(160,50)(190,50) \Text(175,47)[t]{$\tilde{b}$}
  \DashArrowArcn(160,50)(30,180,90){2} \Text(140,80)[br]{$\tilde{\mu}_L$}
  \DashArrowArc(160,50)(30,0,90){2} \Text(180,80)[bl]{$\tilde{e}_{Ri}$}
  \Vertex(160,50){2}  \Vertex(160,80){2}
  \ArrowLine(215,50)(190,50) \Text(205,45)[t]{$e_R$}
  \Photon(192,60)(215,75){2}{3.5} \Text(212,83)[b]{$\gamma$}
  \Text(280,25)[t]{(c)}
  \ArrowLine(225,50)(250,50) \Text(238,45)[t]{$\mu_L$}
  \ArrowLine(265,50)(250,50) \Text(259,47)[t]{$\tilde{h}_d^0$}
  \ArrowLine(265,50)(280,50) \Text(275,47)[t]{$\tilde{h}_u^0$}
  \ArrowLine(295,50)(280,50) \Text(290,47)[t]{$\tilde{b}$}
  \ArrowLine(295,50)(310,50) \Text(303,47)[t]{$\tilde{b}$}
  \Vertex(265,50){2} \Vertex(280,50){2} \Vertex(295,50){2}
  \DashArrowArc(280,50)(30,0,180){2} \Text(280,88)[b]{$\tilde{e}_{Ri}$}
  \ArrowLine(335,50)(310,50)  \Text(325,45)[t]{$e_R$}
  \Photon(312,60)(335,75){2}{3.5} \Text(332,83)[b]{$\gamma$}
  \Text(400,25)[t]{(d)}
  \ArrowLine(345,50)(370,50) \Text(358,45)[t]{$\mu_L$}
  \ArrowLine(400,50)(370,50) \Text(388,47)[t]{$\tilde{h}_d^0$}
  \ArrowLine(400,50)(430,50) \Text(415,47)[t]{$\tilde{b}$} \Vertex(400,50){2}
  \DashArrowArc(400,50)(30,0,180){2} \Text(400,88)[b]{$\tilde{e}_{Ri}$}
  \ArrowLine(455,50)(430,50)  \Text(445,45)[t]{$e_R$}
  \Photon(432,60)(455,75){2}{3.5} \Text(452,83)[b]{$\gamma$}
\end{picture}
\caption{Feynman diagrams contributing to the $\mu \rightarrow e\gamma$ 
 process.  The chirality flips represented by blobs come from the 
 muon mass, the gaugino and Higgsino masses, the left-right slepton 
 mixing, and gaugino-Higgsino mixings.}
\label{fig:mu-e-gamma}
\end{center}
\end{figure}

In this sub-section we consider $\mu \rightarrow e\gamma$ decay. This 
process arises from the one-loop diagrams of Fig.~\ref{fig:mu-e-gamma},
where sleptons and neutralinos circulate in the loop.  We first 
consider the two diagrams of Fig.~\ref{fig:mu-e-gamma}a and 
Fig.~\ref{fig:mu-e-gamma}b, in which both vertices arise from the 
$U(1)_Y$ gauge coupling. In this case, the decay rate for 
$\mu \rightarrow e\gamma$ is given by
\begin{equation}
  \Gamma(\mu \rightarrow e\gamma) = \frac{\alpha}{4}
    m_\mu^3 \left| F^{(a)} + F^{(b)} \right|^2,
\end{equation}
where $\alpha$ and $m_\mu$ represent the fine structure constant and 
the muon mass, respectively. Here, the amplitudes $F^{(a)}$ and 
$F^{(b)}$ are given by
\begin{eqnarray}
  F^{(a)} &=& \frac{\alpha}{4\pi\cos^2\theta_w}m_\mu 
    W^e_{\tau\mu} W^{e*}_{\tau e} 
    \left[ G_1(m_{\tilde{\tau}}^2) - G_1(m_{\tilde{e}}^2) \right],
\\
  F^{(b)} &=& -\frac{\alpha}{4\pi\cos^2\theta_w}m_\mu 
    W^e_{\tau\mu} W^{e*}_{\tau e} 
\nonumber\\
&& \times \left[ (2\tilde{m}+\delta A+\mu\tan\beta) 
       \left\{ G_2(m_{\tilde{l}}^2,m_{\tilde{\tau}}^2) 
              - G_2(m_{\tilde{l}}^2,m_{\tilde{e}}^2) \right\}
       - \tilde{m} G_2(m_{\tilde{l}}^2,m_{\tilde{\tau}}^2) \right],
\end{eqnarray}
where $\theta_w$ is the Weinberg angle, and the functions $G_1$ and 
$G_2$ are defined in Ref.~\cite{Barbieri:1995tw}.  Substituting the 
predictions of the superparticle mass spectrum in our theory, these 
amplitudes are written in terms of a single mass scale $\tilde{m}$ as 
\begin{equation}
  F^{(a)} + F^{(b)} 
    \simeq \left( 0.2 + 0.8\, {\rm sgn}(\mu) \tan\beta \right)
      \frac{\alpha}{4\pi\cos^2\theta_w}
      \frac{m_\mu}{\tilde{m}^2}\, W^e_{\tau\mu} W^{e*}_{\tau e}.
\label{eq:lfv-ampl}
\end{equation}
We find that in the parameter region of our interest, 
$\tan\beta \gsim 3$, the $\mu \rightarrow e\gamma$ process is 
dominated by the contribution proportional to $\tan\beta$, which 
comes from the diagram of Fig.~\ref{fig:mu-e-gamma}b.

The diagrams containing the Yukawa vertex, Fig.~\ref{fig:mu-e-gamma}c 
and Fig.~\ref{fig:mu-e-gamma}d, are also dominated by the contribution 
enhanced by $\tan\beta$.  It comes from the diagram of 
Fig.~\ref{fig:mu-e-gamma}c and destructively interferes with the 
previous contribution in Eq.~(\ref{eq:lfv-ampl}) \cite{Hisano:1996qq}.
Using our superparticle spectrum, we find $F^{(c)} \simeq -0.4 F^{(b)}$, 
which reduces the rate of $\mu \rightarrow e\gamma$ by about a factor 
of 3 compared with that given by Eq.~(\ref{eq:lfv-ampl}).  Therefore, 
the $\mu \rightarrow e\gamma$ decay rate in our theory is given by 
\begin{equation}
  \Gamma(\mu \rightarrow e\gamma) 
    \simeq \frac{c_\mu^2 \alpha^3}{64\pi^2\cos^4\theta_w}\,
      \frac{m_\mu^5}{\tilde{m}^4}\, |W^e_{\tau\mu}|^2 
      |W^e_{\tau e}|^2 \tan^2\beta,
\end{equation}
where $c_\mu \simeq 0.5$.  Dividing this by the total decay rate of the 
muon, $\Gamma(\mu) \simeq \Gamma(\mu \rightarrow e\nu\bar{\nu}) \simeq 
(\alpha^2/384\pi\sin^4\theta_w\cos^4\theta_w)(m_\mu^5/M_Z^4)$, 
we finally obtain the branching ratio for the $\mu \rightarrow e\gamma$ 
decay in our theory
\begin{equation}
  {\rm Br}(\mu \rightarrow e\gamma) 
    \simeq 3 \times 10^{-11} 
      \left(\frac{200~{\rm GeV}}{\tilde{m}}\right)^4
      \left(\frac{|W^e_{\tau\mu}|}{0.04}\right)^2 
      \left(\frac{|W^e_{\tau e}|}{0.01}\right)^2 
      \left(\frac{\tan\beta}{5.0}\right)^2.
\label{eq:br-meg}
\end{equation}
Here, we have normalized elements of the new mixing matrix $W^e$ by 
the corresponding values in the CKM matrix.  This is well motivated 
because $W^e$ comes from a rotation of the right-handed charged leptons 
$e$, and the rotation of $e$ is expected to be similar to that of the 
left-handed quarks $q$, which determines the CKM matrix.

The prediction given in Eq.~(\ref{eq:br-meg}) is very interesting, 
since it gives a number close to the present experimental bound 
${\rm Br}(\mu \rightarrow e\gamma) \lsim 1.2 \times 10^{-11}$ 
\cite{Brooks:1999pu}.  While we expect an uncertainty of a factor of 
a few in the estimate of Eq.~(\ref{eq:br-meg}), for example from 
uncertainties for the value of $m_{\tilde{\tau}}^2$ discussed in 
sub-section \ref{subsec:theory2}, we can still say that the present 
$\mu \rightarrow e\gamma$ decay experiment has already probed our 
theory up to about $\tilde{m} \simeq 200~{\rm GeV}$ ($300~{\rm GeV}$) 
for $\tan\beta = 5$ ($10$).\footnote{
Even if additional contributions to $m_{\tilde{\tau}}^2$, for example 
from a large $U(1)_X$ gauge coupling, happened to give a cancellation 
between the two diagrams of Fig.~\ref{fig:mu-e-gamma}b and 
Fig.~\ref{fig:mu-e-gamma}c, reducing ${\rm Br}(\mu \rightarrow e\gamma)$ 
by a few orders of magnitude, the $\mu$-$e$ conversion in nuclei, 
discussed in the next sub-section, will provide an effective probe 
of the theory \cite{Hisano:1996qq}.}
Furthermore, a new experiment is under construction at PSI which aims 
for a sensitivity of ${\rm Br}(\mu \rightarrow e\gamma)$ at a 
$10^{-14}$ level \cite{PSI}.  Since ${\rm Br}(\mu \rightarrow e\gamma) 
\lsim 10^{-14}$ corresponds to $\tilde{m} \gsim 1.5~{\rm TeV}$ 
($2~{\rm TeV}$) for $\tan\beta = 5$ ($10$) in Eq.~(\ref{eq:br-meg}), 
it will probe essentially all the parameter region of our theory where 
radiative electroweak symmetry breaking occurs naturally.

We now discuss the effect of a possible mass splitting between 
$\tilde{l}_{1,2}$ and $\tilde{l}_3$.  In deriving a general expression 
for the lepton flavor violation in Eq.~(\ref{eq:lfvwe}), we have 
used the fact that the masses for $\tilde{l}_i$ are all degenerate 
in our theory, $m_{\tilde{l}_{1,2}}^2 = m_{\tilde{l}_3}^2$.  However, 
this relation is not strictly correct; in particular, if $\tan\beta$ 
becomes larger, $m_{\tilde{l}_3}^2$ becomes smaller than 
$m_{\tilde{l}_{1,2}}^2$ due to the renormalization group effect 
through the tau Yukawa coupling.  The non-degeneracy between 
$m_{\tilde{l}_{1,2}}^2$ and $m_{\tilde{l}_3}^2$ introduces new lepton 
flavor mixing angles in addition to those in $W^e$, and causes new 
diagrams for the $\mu \rightarrow e\gamma$ process.\footnote{
The situation is somewhat similar to lepton flavor violation in 4D 
supersymmetric $SO(10)$ theories, where both $\tilde{l}$ and $\tilde{e}$ 
non-degeneracies arising from radiative corrections involving the large 
Yukawa coupling for the third generation \cite{Barbieri:1995tw}.}
Since the amplitudes of these new diagrams can be proportional to 
$m_\tau$ rather than $m_\mu$, they might dominate the contribution 
from the diagrams of Fig.~\ref{fig:mu-e-gamma}, when $\tan\beta$ is 
large. The relative size of this new contribution to the previous 
one is expected to be of order $(m_\tau/m_\mu) \delta_{\tilde{l}}$ 
in the amplitudes, where $\delta_{\tilde{l}} \equiv 
(m_{\tilde{l}_{1,2}}^2-m_{\tilde{l}_3}^2)/m_{\tilde{l}_{1,2}}^2$. 
There may also be an additional suppression by a factor of $\epsilon$ 
if the charged lepton mass matrix is given by the form of 
Eq.~(\ref{eq:yde}), because the relevant rotation angles for $l_i$ are 
then of order $\epsilon^2$ while those for $e_i$ are of order 
$\epsilon$. In any event, since $\delta_{\tilde{l}} \simeq 0.02$ 
($0.05$) for $\tan\beta=10$ ($15$), we find that the contribution from 
$\delta_{\tilde{l}} \neq 0$ can be safely neglected in the parameter 
region of our interest, $\tan\beta \lsim 15$, and the branching ratio 
of $\mu \rightarrow e\gamma$ is still given by Eq.~(\ref{eq:br-meg}), 
which is described by the single mixing matrix $W^e$.
Potentially, the presence of Dirac neutrino Yukawa couplings at 
high energies could also introduce a splitting between 
$m_{\tilde{l}_{1,2}}^2$ and $m_{\tilde{l}_3}^2$, enhancing the decay 
rate of $\mu \rightarrow e\gamma$ \cite{Borzumati:1986qx}.  We 
consider an explicit model of neutrino mass generation in sub-section 
\ref{subsec:neutrino}, where the neutrino Dirac Yukawa couplings are 
suppressed by the volume of the extra dimension and their effects on 
the lepton flavor violating processes are sufficiently small.

The above discussion also tells us something about the location of 
matter in the extra dimension.  Suppose we put the first two generation 
$\bar{\bf 5}$'s, $F_{1,2}$, in the 5D bulk rather than on the $y=0$ 
brane.  In this case, $\tilde{l}_{1,2}$ and $\tilde{l}_3$ have a 
mass splitting of $\delta_{\tilde{l}} = O(1)$ at tree level, so that 
we cannot go to a basis like Eq.~(\ref{eq:yde}) without introducing 
a flavor violation in the slepton mass-squared matrix.  Then, from 
the structure of the charged lepton Yukawa matrix, we find that the 
relevant rotation angles for $l_i$ are of order $\epsilon$ and the 
same order with those for $e_i$.  This means that the $\mu \rightarrow 
e\gamma$ decay rate in the bulk $F_{1,2}$ theory is enhanced by 
a factor of $(m_\tau/m_\mu)^2$ compared with the case of the brane 
$F_{1,2}$ theory.  Therefore, to evade an experimental constraint, the 
overall mass scale for the superparticles, $\tilde{m}$, in the bulk 
$F_{1,2}$ theory must be a factor of $(m_\tau/m_\mu)^{1/2} \simeq 4$ 
higher than that of the brane $F_{1,2}$ theory.  Since the higher 
value of $\tilde{m}$ requires a fine tuning for electroweak symmetry 
breaking, the case of $F_{1,2}$ having a different location than $F_3$ 
($m_{\tilde{l}_{1,2}} \neq m_{\tilde{l}_3}$) is strongly disfavored, 
explaining the unique choice for the matter location we made in 
sub-section \ref{subsec:theory2}.

\subsubsection{Other lepton flavor violating processes}
\label{subsubsec:otherlepton-flavor}

In this sub-section we discuss other lepton flavor violating processes. 
We first consider the branching ratio for the $\mu \rightarrow 3e$ decay. 
This process is dominated by the contribution from photon penguin 
diagrams, since they are enhanced by the phase space integral.  Thus 
there is a simple relation between ${\rm Br}(\mu \rightarrow 3e)$ and 
${\rm Br}(\mu \rightarrow e\gamma)$.  Using Eq.~(\ref{eq:br-meg}), 
we obtain 
\begin{eqnarray}
  {\rm Br}(\mu \rightarrow 3e) 
  &\simeq& \frac{\alpha}{3\pi} 
    \left( \ln\frac{m_\mu^2}{m_e^2} - \frac{11}{4} \right)
    {\rm Br}(\mu \rightarrow e\gamma) 
\nonumber\\
  &\simeq& 2 \times 10^{-13} 
      \left(\frac{200~{\rm GeV}}{\tilde{m}}\right)^4
      \left(\frac{|W^e_{\tau\mu}|}{0.04}\right)^2 
      \left(\frac{|W^e_{\tau e}|}{0.01}\right)^2 
      \left(\frac{\tan\beta}{5.0}\right)^2.
\label{eq:br-me3}
\end{eqnarray}
The present experimental bound is ${\rm Br}(\mu \rightarrow 3e) 
\lsim 1.0 \times 10^{-12}$ \cite{Bellgardt:1987du}.

We next consider $\mu \rightarrow e$ conversion in nuclei.
The conversion rate is well approximated by 
\begin{equation}
  \Gamma(\mu \rightarrow e; X)
  \simeq 16 \alpha^4 Z_{\rm eff}^4 Z |F(q)|^2 
    \Gamma(\mu \rightarrow e\gamma), 
\end{equation}
in the parameter region we are considering.  (The approximation is 
better for larger values of $\tan\beta$.)  Here, $Z$ represents the 
proton number of the nucleus $X$; $Z_{\rm eff}$ and $F(q)$ are the 
effective charge and the nuclear form factor at the momentum transfer 
$q$, respectively.  In the case of $X = {}^{48}_{22}{\rm Ti}$, 
for which $Z=22$, $Z_{\rm eff} \simeq 17.6$ and $|F(q)| \simeq 0.54$ 
\cite{Bernabeu:1993ta}, we obtain the prediction for the 
$\mu \rightarrow e$ conversion rate, normalized by the muon capture 
rate $\Gamma(\mu \rightarrow {\rm capture}; {}^{48}_{22}{\rm Ti}) = 
(2.590 \pm 0.012) \times 10^6~{\rm sec}^{-1}$ \cite{Suzuki:1987jf}, of
\begin{eqnarray}
  {\rm Cr}(\mu \rightarrow e; {}^{48}_{22}{\rm Ti}) 
  &\equiv& \frac{\Gamma(\mu \rightarrow e; {}^{48}_{22}{\rm Ti})}
    {\Gamma(\mu \rightarrow {\rm capture}; {}^{48}_{22}{\rm Ti})} 
\nonumber\\
  &\simeq& 2 \times 10^{-13} 
      \left(\frac{200~{\rm GeV}}{\tilde{m}}\right)^4
      \left(\frac{|W^e_{\tau\mu}|}{0.04}\right)^2 
      \left(\frac{|W^e_{\tau e}|}{0.01}\right)^2 
      \left(\frac{\tan\beta}{5.0}\right)^2.
\label{eq:br-mec}
\end{eqnarray}
The present experimental bound is ${\rm Cr}(\mu \rightarrow e; 
{}^{48}_{22}{\rm Ti}) \lsim 4.3 \times 10^{-12}$ \cite{Dohmen:mp}.
It is interesting to note that future experiments may probe 
$\mu \rightarrow e$ conversion in nuclei with a sensitivity below 
$10^{-16}$ \cite{BNL}.

Another important lepton flavor violating process is $\tau \rightarrow 
\mu\gamma$ decay.  The branching ratio for $\tau \rightarrow \mu\gamma$ 
is related to that of $\mu \rightarrow e\gamma$ by 
\begin{equation}
  \frac{\Gamma(\tau \rightarrow \mu\gamma)}{\Gamma(\mu \rightarrow e\gamma)}
  = \left| \frac{W^e_{\tau\tau}}{W^e_{\tau e}} \right|^2 
    {\rm Br}(\tau \rightarrow \mu\nu\bar{\nu}).
\end{equation}
Since ${\rm Br}(\tau \rightarrow \mu\nu\bar{\nu}) = 
(17.37 \pm 0.07)\%$, we obtain
\begin{equation}
  {\rm Br}(\tau \rightarrow \mu\gamma) 
    \simeq 5 \times 10^{-8} 
      \left(\frac{200~{\rm GeV}}{\tilde{m}}\right)^4
      \left(\frac{|W^e_{\tau\mu}|}{0.04}\right)^2 
      \left(\frac{|W^e_{\tau\tau}|}{1.0}\right)^2 
      \left(\frac{\tan\beta}{5.0}\right)^2.
\label{eq:br-tmg}
\end{equation}
The present experimental bound comes from CLEO: ${\rm Br}(\tau 
\rightarrow \mu\gamma) \lsim 1.1 \times 10^{-6}$ \cite{Ahmed:1999gh}.
The $B$ factories at KEK and SLAC will improve the bound to the level 
of $10^{-7}$.  Note that the combination of lepton flavor violation 
mixing angles, $\theta^e_{ij}$, appearing in Eq.~(\ref{eq:br-tmg}) 
is different from that in Eqs.~(\ref{eq:br-meg}, \ref{eq:br-me3}, 
\ref{eq:br-mec}).  Therefore, in principle, 
we can determine all the lepton flavor violation mixing angles, 
$\theta^e_{12}$ and $\theta^e_{23}$, by measuring both 
$\mu \rightarrow e$ and $\tau \rightarrow \mu$ transition rates, 
if we know $\tilde{m}$ and $\tan\beta$ from independent measurements 
of the superparticle spectrum.

\subsection{Hadronic flavor violation}
\label{subsec:hadron-flavor}

The structure of hadronic flavor violation in our theory is much more 
complicated than that of lepton flavor violation.  In addition to the 
complication arising from the CKM matrix, there is also a 
complication coming from the fact that up-type squarks have flavor 
non-universalities in both the left-handed and right-handed 
mass-squared matrices. Although it is still possible to derive a 
general Lagrangian for hadronic flavor violation, as in the leptonic 
case of Eq.~(\ref{eq:lfvwe}), the resulting expression is not 
particularly illuminating, so that here we focus on estimating 
constraints on the overall superparticle mass scale, $\tilde{m}$, 
coming from various hadronic flavor violating processes.

We start with the $b \rightarrow s\gamma$ decay process. In the 
minimal supergravity scenario of supersymmetry breaking, it is often 
claimed that the exact $b/\tau$ unification is consistent with the 
constraint from the $b \rightarrow s\gamma$ decay only when the 
squark masses are rather large.  This is because the exact $b/\tau$ 
unification requires the negative sign of $\mu$ and/or large values 
for $\tan\beta$, which enhances the supersymmetric contribution to 
the $b \rightarrow s\gamma$ process.  In our theory, however, there 
is a correction from the unified scale, Eq.~(\ref{eq:delta-mb-Mc}), 
allowing us to consider the parameter region of $\tan\beta \simeq 
5{\rm -}10$. Thus the squark masses do not have to be very large.  
Furthermore, the $b \rightarrow s\gamma$ decay is a rigid constraint 
on the parameter space only if squark mass matrices are diagonal in 
the super-CKM basis.  This is not the case in our theory, since the 
squark mass matrices have flavor non-universalities at the 
compactification scale.  By rotating to the super-CKM basis, this 
introduces flavor off-diagonal elements in the squark mass matrices, 
which are not determined by the observed CKM mixing angles. Therefore, 
in our theory we cannot extract a definite bound from the 
$b \rightarrow s\gamma$ process.  A rough bound is obtained as 
$\tilde{m} \gsim 100\sqrt{\tan\beta}~{\rm GeV}$ by estimating the 
size of the chargino exchange diagrams.

We next consider constraints from neutral meson mixings. 
Since flavor violation in the squark sector is relatively small due to 
the gluino focusing effect, we can use the mass insertion approximation 
to estimate flavor violating processes.  The relevant quantities are 
the mass insertion parameters $\delta_{ij}$ defined by 
$(\delta^f_{ij})_{XY} \equiv (m_{\tilde{f}}^2)_{iX,jY}
/\sqrt{(m_{\tilde{f}}^2)_{iX,iX}(m_{\tilde{f}}^2)_{jY,jY}}$ 
in the super-CKM basis, where $f=u,d$ specifies the up-type or 
down-type sector and $X,Y=L,R$ the left-handed or right-handed 
chirality; $i,j=1,2,3$ are the generation indices.
The $K_L$-$K_R$ mass difference, $\Delta m_K$, gives constraints on 
$(\delta^d_{12})_{XY}$.  In our theory, these quantities are estimated 
as $(\delta^d_{12})_{LL} \approx (m_{\tilde{q}_B}^2-m_{\tilde{q}_3}^2)
V^{\rm CKM}_{td}V^{\rm CKM}_{ts}/m_{\tilde{q}_B}^2 \approx 10^{-4}$, 
$(\delta^d_{12})_{RR} \approx 0$, and $(\delta^d_{12})_{LR}, 
(\delta^d_{12})_{RL} \lsim m_s\mu\tan\beta/m_{\tilde{q}_B}^2 
\approx 10^{-4}\tan\beta (200~{\rm GeV}/\tilde{m})$, assuming no 
contribution from brane-localized kinetic terms.  Therefore, we find 
that the supersymmetric contribution to $\Delta m_K$ is negligible 
for $\tilde{m} \gsim 200~{\rm GeV}$ \cite{Gabbiani:1996hi}. Since the 
constraint from the $CP$ violating parameter $\varepsilon_K$ is 
somewhat stronger than that from $\Delta m_K$, however, there may be 
an observable effect on $\varepsilon_K$ in the case of lower values 
for $\tilde{m}$ (and larger $\tan\beta$) if the relevant phase is of 
order unity.  Analogous considerations can also be made for 
$B$-$\bar{B}$ and $D$-$\bar{D}$ mixings, which constrain 
$(\delta^d_{13})_{XY}$ and $(\delta^u_{12})_{XY}$, respectively.  
Again, we obtain essentially no constraint for 
$\tilde{m} \gsim 200~{\rm GeV}$.

\section{A Realistic Completion of the Theory}
\label{sec:develop}

In this section we provide a realistic extension of our theory, 
incorporating small neutrino masses and a natural generation of the 
supersymmetric mass term ($\mu$ term) and the holomorphic supersymmetry 
breaking mass term ($\mu B$ term) for the Higgs doublets.  In our 
example, these two issues are related through a single dynamics 
triggering the spontaneous breaking of the $U(1)_X$ gauge symmetry.
Small neutrino masses are generated by integrating out right-handed 
neutrino superfields through the see-saw mechanism \cite{Seesaw}. 
The $\mu$ and $\mu B$ terms of the correct size are generated by the 
vacuum readjustment mechanism of Ref.~\cite{Hall:2002up}. 
The generated $B$ parameter is real, so that there is no 
supersymmetric $CP$ problem.  

\subsection{Neutrino masses and $U(1)_X$ gauge interaction}
\label{subsec:neutrino}

Recent neutrino experiments have provided strong evidence that 
neutrinos have small masses and the different flavors are mixed 
\cite{Fukuda:1998mi, Ahmad:2001an}.  In our theory small neutrino 
masses can be generated through the see-saw mechanism by 
introducing three generations of right-handed neutrino superfields. 
They could be either brane fields, $N$, or bulk fields, 
$\{ N,N^c \}$ with $\eta_N=1$.  In both cases, the Yukawa couplings 
and Majorana masses for $N$ are located on the $y=0$ brane
\begin{equation}
  S = \int d^4x \; dy \; \delta(y)
    \biggl[ \int d^2\theta \Bigl( y_N F N H 
	+ \frac{\kappa_R}{2} N N \Bigr) + {\rm h.c.} \biggr].
\label{eq:nu-yukawa}
\end{equation}
The $U(1)_R$ charges for the right-handed neutrinos are given by 
$R(N) = R(N^c) = 1$, as for the other matter fields $T$ and $F$, so 
that the above superpotential terms are invariant under $U(1)_R$.  
After integrating over the extra dimension, these terms give the 4D 
neutrino Dirac Yukawa couplings, $[y_\nu L N H_u]_{\theta^2}$, and the 
4D Majorana mass terms $[(M_R/2) N N]_{\theta^2}$.  Thus, integrating 
out the right-handed neutrinos we obtain the operators 
$[(y_\nu^2/2M_R) L L H H]_{\theta^2}$ at low energies, which provide 
small masses $m_\nu \sim y_\nu^2 v^2/M_R$ to the observed 
(left-handed) neutrinos.

What is the scale for $M_R$?  It depends on the size of the 4D neutrino 
Dirac Yukawa couplings, $y_\nu$.  If $N$ is the brane field, we expect 
that $y_\nu = O(1)$ at least for the third generation, so that 
$M_R$ could be as high as the compactification scale, $M'_c$, keeping 
$m_\nu \simeq 0.03{\rm -}0.1~{\rm eV}$ suggested from the atmospheric 
neutrino oscillation data.  Thus, in this case, the interactions in 
Eq.~(\ref{eq:nu-yukawa}) will not affect the superparticle mass spectrum 
much, since the superparticle masses are effectively generated at $M'_c$.
On the other hand, if $N$ is in the bulk, the 4D Yukawa coupling 
$y_\nu$ is volume suppressed, so that $M_R$ will be smaller than $M'_c$ 
to obtain desired values for the neutrino masses.  Therefore, the 
neutrino interactions in Eq.~(\ref{eq:nu-yukawa}) could potentially 
affect the superparticle mass spectrum through renormalization group 
evolutions.  Below, we will choose $N$ to be in the bulk and give 
an example of the complete theory, but the same mechanism works also 
for the brane $N$ case with obvious modifications of the statements 
that are specific to the bulk $N$ case.

We introduce a $U(1)_X$ gauge interaction ($\subset SO(10)/SU(5)$) in 
the bulk, under which matter and Higgs fields transform as $T(1)$, 
$F(-3)$, $N(5)$, $H(-2)$ and $\bar{H}(2)$ (for the bulk fields, the 
conjugated chiral superfields have opposite quantum numbers to the 
corresponding unconjugated fields).  Then, the Majorana masses for $N$ 
are generated by the spontaneous breaking of the $U(1)_X$ symmetry, 
explaining why they take values smaller than the value expected from 
pure dimensional analysis.  One way of realizing this is to 
introduce $U(1)_X$ breaking fields $\Psi(10)$ and $\bar{\Psi}(-10)$ 
with the following superpotential
\begin{equation}
  S = \int d^4x \; dy \; \delta(y)
    \biggl[ \int d^2\theta \Bigl( f X (a \Psi\bar{\Psi} - \Lambda^2) 
    + \frac{\kappa}{2} \bar{\Psi} N N \Bigr) + {\rm h.c.} \biggr],
\label{eq:MR-gen}
\end{equation}
where $X$ is a gauge singlet field, and $\Lambda$ is a scale 
arising from the dynamics of some strongly coupled gauge 
interaction.\footnote{An explicit example for the strongly 
interacting sector is given in Ref.~\cite{Hall:2002up}, where the 
$SU(2)_S$ gauge interaction with four doublet fields ${\cal Q}_i$ 
and five singlet fields $X^a$ having an appropriate superpotential 
is considered.  We here assume that this strong sector is localized 
on the $y=0$ brane, which explains why this gauge interaction 
is stronger than the other gauge interactions such as the 
standard model ones and $U(1)_X$.} 
This superpotential forces vacuum expectation values for the 
$\Psi$ and $\bar{\Psi}$ fields, $\vev{\Psi} = \vev{\bar{\Psi}} = 
\Lambda/\sqrt{a}$, giving Majorana masses for the right-handed 
neutrinos, $\kappa_R = \kappa\Lambda/\sqrt{a}$.  Thus the superpotential 
Eq.~(\ref{eq:MR-gen}) effectively reproduces the second term of the 
superpotential Eq.~(\ref{eq:nu-yukawa}).  Note that the whole sector 
of neutrino mass generation is invariant under $U(1)_R$.  
Specifically, the $U(1)_R$ charge assignment for various fields is 
given by $R(X) = 2$ and $R(\Psi) = R(\bar{\Psi}) = 0$, and $U(1)_R$ 
is not broken by the dynamics of this sector.\footnote{
In the $SU(2)_S$ example of Ref.~\cite{Hall:2002up}, $R(X^a) = 2$ 
and $R({\cal Q}_i) = 0$ so that $U(1)_R$ does not have an anomaly 
for $SU(2)_S$ (i.e. $R(\Lambda) = 0$).}

We finally comment on the possible effect of the interactions in 
Eq.~(\ref{eq:nu-yukawa}) on the superparticle spectrum.  In the case 
that $N$ propagates in the bulk, the 4D neutrino Yukawa coupling 
$y_\nu$ receives a volume suppression and is expected to be $y_\nu 
\lsim \epsilon \approx 0.1$.  Therefore, the effect of this coupling 
on the low energy spectrum will be small.  In particular, the 
splitting between $m_{\tilde{l}_{1,2}}$ and $m_{\tilde{l}_{3}}$ caused 
by this coupling through the renormalization group evolution is 
expected to be sufficiently small that the previous estimates for 
lepton flavor violating processes will remain intact.  On the other 
hand, smaller Yukawa couplings suggest a smaller scale for the 
Majorana mass $M_R$.  However, this does not necessarily mean that 
the scale for the $U(1)_X$ breaking is small.  In fact, by tracing 
the volume suppression factors, we find that the vacuum expectation 
values for $\Psi$ and $\bar{\Psi}$ are not much smaller than the 
compactification scale.  Therefore, the radiative contribution to 
the stau mass from the $U(1)_X$ gaugino is also expected to be 
small.\footnote{
Such expectations are not definitive: there is a possibility that 
this correction is sizable due to longer running distances and/or 
larger values of the $U(1)_X$ gauge coupling.}

\subsection{Origin of $\mu$ term}
\label{subsec:mu}

The neutrino mass generation of the previous sub-section also provides 
a natural mechanism for generating the $\mu$ term \cite{Hall:2002up}. 
One easy way to implement this mechanism is to put the $\Psi$ 
and $\bar{\Psi}$ fields in the 5D bulk as two hypermultiplets, 
$\{ \Psi,\Psi^c \} + \{ \bar{\Psi},\bar{\Psi}^c \}$ with $\eta_{\Psi} 
= \eta_{\bar{\Psi}} = 1$, having the superpotential coupling 
Eq.~(\ref{eq:MR-gen}) on the $y=0$ brane.  Now, suppose that the scale 
of the vacuum expectation values for $\Psi$ and $\bar{\Psi}$ is lower 
than the compactification scale. This occurs in some parameter region 
of the theory, for example, if the dimensionless coefficients of the 
neutrino Yukawa couplings are somewhat smaller than the other couplings. 
In this case, the 4D effective theory below $M'_c$ contains holomorphic 
supersymmetry breaking terms in addition to the supersymmetric terms 
arising from Eq.~(\ref{eq:MR-gen}).  Using the general result of 
Eq.~(\ref{eq:soft}), we find that holomorphic supersymmetry breaking 
parameters for $X\Psi\bar{\Psi}$ and $X\Lambda^2$ terms are different 
($-2\tilde{m}$ and $0$ in the case that $X$ is a brane field).
Then, by minimizing the scalar potential, we find that the $X$ 
superfield develops vacuum expectation values of order the weak scale 
in both lowest and highest components, $\vev{X} \sim \tilde{m}$ and 
$\vev{F_X} \sim \tilde{m}^2$.  Therefore, by introducing the coupling
\begin{equation}
  S = \int d^4x \; dy \; \delta(y)
    \biggl[ \int d^2\theta \lambda X H \bar{H} + {\rm h.c.} \biggr],
\label{eq:mu-gen}
\end{equation}
we obtain the $\mu$ and $\mu B$ terms of the correct size. Since the 
$B$ parameter generated through this mechanism is real, there is no 
supersymmetric $CP$ problem.  This means that the $\mu$ parameter is 
real in our phase convention where $\tan\beta$ (and thus the $\mu B$ 
parameter) is taken to be real.  Note that the above coupling 
in Eq.~(\ref{eq:mu-gen}) respects the $U(1)_R$ symmetry, so that the 
whole system is still $U(1)_R$ invariant.  The $U(1)_R$ breaking lies 
only in the supersymmetry breaking terms arising from the boundary 
conditions, which can be viewed as a vacuum expectation value 
for an auxiliary field in the 5D gravity multiplet.

Finally, we briefly comment on an alternative possibility of generating 
the $\mu$ term.  Instead of relying on the above vacuum readjustment 
mechanism, we could introduce the singlet field $S$ at the weak scale 
and write down the superpotential terms 
$[S H \bar{H} + S^3]_{\theta^2}$ on the $y=0$ brane.  Then, in some 
parameter region, the lowest and highest components of the $S$ 
field get vacuum expectation values of the order of the weak 
scale, generating $\mu$ and $\mu B$ terms of the correct order 
\cite{Nilles:1982dy}. In this case the $U(1)_R$ symmetry is explicitly 
broken to the discrete $Z_{4,R}$ subgroup by the $S^3$ term in the 
superpotential, but it is still sufficient to suppress unwanted terms 
such as tree-level $\mu$ term and dimension four and five proton 
decay operators.  Since the Higgs quartic couplings receive an 
additional contribution from the superpotential term $S^3$, 
the physical Higgs boson mass can be larger than that in the MSSM-type 
models where there is no singlet field around the weak scale.

\section{Alternative Possibility}
\label{sec:alt}

We have seen that the location of matter in our theory is uniquely 
determined as a consequence of the ``large'' extra dimension 
and by the requirements of $b/\tau$ unification and naturalness for 
electroweak symmetry breaking.  The predictive framework for gauge 
coupling unification requires strong coupling at the cutoff scale, 
and thus the large volume for the extra dimension.  The location 
for $T_1$ and $T_3$ are then determined to be the bulk and the brane 
by considering the constraint from dimension six proton decay and 
the size of the top Yukawa coupling, respectively.  Breaking 
supersymmetry by boundary conditions, the first two generations 
having the same gauge quantum numbers must be located in the same 
place to evade constraints from flavor changing neutral current 
processes; hence $T_2$ must be located in the bulk.
The $b/\tau$ unification requires $F_3$ on the brane, and finally 
$F_{1,2}$ are located in the same place as $F_3$ to avoid too large 
lepton flavor violating processes. 

Obviously, relaxing some of these requirements allow us to consider 
other possibilities for the matter location, which we explore in this 
section in the framework of KK grand unification with 
boundary condition supersymmetry breaking.  We first observe that 
if we insist on the predictive scheme for gauge coupling unification, 
namely the strong coupling scenario, the location of $T_i$ are 
completely fixed: $T_3$ on the brane and $T_{1,2}$ in the bulk. 
Then we find that all $F_i$ must be put together in the same place to 
evade excessive lepton flavor violating processes which would push up 
the overall mass scale for superpartners.  Therefore, we have only two 
choices for the location of $F_i$: all $F_i$ on the brane, which we 
have adopted so far in this paper, or all $F_i$ in the bulk.  In the 
latter case of $F_i$ in the bulk, there is no $b/\tau$ unification 
because the bottom and tau Yukawa couplings come from different 
interactions that are not related by the $SU(5)$ symmetry.
However, we instead obtain an understanding of the $t/b$ mass ratio, 
since $TF\bar{H}$ type Yukawa couplings are now suppressed by the 
volume factor compared with $TTH$ type couplings.  Thus we find that 
the case of bulk $F_i$ is also interesting, especially if future 
improvements of extracting $m_b$ from data determine $m_b$ to be in 
the lower part of the presently allowed experimental region.

The superparticle spectrum in the case of bulk $F_i$ is different 
from that of the brane $F_i$ case, corresponding to the difference of 
soft supersymmetry breaking parameters at the compactification scale.
In particular, we now have to use the values of $\tilde{d}_B$ and 
$\tilde{l}_B$ in Table~\ref{table:soft} for the first two generation 
$\tilde{d}_{1,2}$ and $\tilde{l}_{1,2}$, instead of $\tilde{d}_b$ 
and $\tilde{l}_b$.  Similarly, the values for the third generation 
squark and slepton masses are changed from 
$(\tilde{q}_3,\tilde{u}_3,\tilde{d}_3,\tilde{l}_3,\tilde{e}_3)
\simeq (390,310,420,140,66)$ to $(390,310,460,240,63)$ in 
Table~\ref{table:soft}.  The sizes for the $A$ terms are also changed: 
$(A_t,A_b,A_\tau) \simeq (-410,-730,-320) \rightarrow (-410,-930,-520)$.
These changes affect the expression of the weak scale threshold 
correction to gauge coupling unification.  The new expression is 
given by $\delta\alpha_s|_{\rm susy}^{\rm KK} \simeq 0.0034 - 0.0030 
\ln(\tilde{m}/M_Z)$ instead of Eq.~(\ref{eq:assusykk}), which makes 
the prediction for the QCD coupling slightly lower than the brane 
$F_i$ case for the same value of $\tilde{m}$.  Note also that the bulk 
$F_i$ theory allows both signs for the $\mu$ parameter, since the 
supersymmetric threshold correction to $m_b$ can now have either sign.

Proton decay in the bulk $F_i$ theory is very much suppressed.
The decay by flavor mixings receives the suppression of order 
$(V_{13}V_{32})^2$ in the amplitude, giving the lifetime estimate 
$\tau(p \rightarrow \mu^+ K^0) \approx 10^{44}~{\rm years}$.
The decay through brane kinetic operators is also suppressed: 
$(M'_c/M_s)^2$ suppression in the amplitude, giving 
$\tau(p \rightarrow e^+ \pi^0, \cdots) \approx 10^{39}~{\rm years}$.

The rates for lepton flavor violating processes are also subject to 
important changes.  We first consider the contributions to the $\mu 
\rightarrow e\gamma$ process proportional to $\tan\beta$, which 
come from the diagrams of Fig.~\ref{fig:mu-e-gamma}b and 
Fig.~\ref{fig:mu-e-gamma}c.  In the theory where $F_i$ are on the brane, 
the diagram with the Yukawa vertex, Fig.~\ref{fig:mu-e-gamma}c, is only 
$40\%$ of the pure gauge diagram, Fig.~\ref{fig:mu-e-gamma}b, in the 
amplitude.  However, in the theory with bulk $F_i$, the left-handed 
sleptons are heavier, making the contribution from the diagram of 
Fig.~\ref{fig:mu-e-gamma}b smaller.  By explicitly calculating the 
two diagrams, we find that the two contributions have a comparable 
size with the opposite sign.  Hence the potentially leading contribution 
with the $\tan\beta$ enhancement turns out to be small due to the 
cancellation between the two diagrams.  The precise calculation for 
the value of the remaining contribution is difficult without a precise 
knowledge of the superparticle mass spectrum, but we can roughly 
estimate the expected size for ${\rm Br}(\mu \rightarrow e\gamma)$ 
by evaluating the piece which is not proportional to $\tan\beta$. 
This piece is larger than the previous case because the $A$ terms are 
larger due to the bulk location of $F_i$; specifically, the coefficient 
in Eq.~(\ref{eq:lfv-ampl}) becomes $0.2 \rightarrow 1$.  Overall, we 
expect lepton flavor violation rates in the bulk $F_i$ theory 
to be one or two orders of magnitude smaller than the corresponding 
ones in the brane $F_i$ theory.  However, since the parameter region 
which leads to the exact cancellation is different for different 
processes, for instance between the $\mu \rightarrow e\gamma$ decay 
and the $\mu$-$e$ conversion in nuclei, some of them will survive 
the strong cancellation and could still have comparable rates to 
the case of the brane $F_i$ theory.

We finally comment on the possibility of relaxing the strong coupling 
assumption at $M_s$.  In this case the volume of the extra dimension 
does not necessarily have to be very large, so we can consider yet 
other possibilities for the matter location, although we loose the high 
predictivity for gauge coupling unification leading to the prediction 
as Eq.~(\ref{eq:askk}).  In particular, the location of $T_i$ is now 
not completely determined.  The cases where all matter fields are 
located on the brane or in the bulk have been discussed in 
Ref.~\cite{Barbieri:2001yz}.

\section{Conclusions}
\label{sec:concl}

The merging of gauge couplings at energies of order $10^{16}~{\rm GeV}$ 
heralds some new unified physics beyond the supersymmetric desert. 
An attractive new option for this physics is that extra dimensions 
of spacetime are resolved, with local defects explicitly breaking the 
unified gauge symmetry \cite{Hall:2001pg, Hall:2001xb}. Advances of 
4D grand unification are kept, while many of the problems are overcome. 
An understanding of quark and lepton quantum numbers is preserved, 
while gauge coupling unification emerges in the limit that the defects 
are embedded in a large bulk. The breaking of gauge symmetry is 
automatic, as is a large mass gap between light and heavy gauge fields 
and between doublet and triplet Higgs fields \cite{Kawamura:2001ev}. 
Proton stability from operators of dimension four and five is guaranteed 
by a continuous $R$ symmetry of the underlying 5D supersymmetric theory, 
while quark-lepton mass relations are only expected for heavy 
generations. A crucial question for this new framework is: how can 
it be tested?

In this paper we have developed the minimal KK grand unified theory 
of Ref.~\cite{Hall:2001xb}, based on $SU(5)$ gauge interactions 
in 5D, into a complete, realistic theory. The major new ingredient 
is to break supersymmetry by boundary conditions applied to the same 
fifth dimension that breaks the gauge symmetry. If this is the correct 
effective theory of nature, over the next decade experiments will 
provide convincing evidence for it, measuring the locations of quarks 
and leptons in the bulk. While it seems to us a natural way to 
incorporate supersymmetry breaking into KK unified theories, there 
are clearly other possibilities, which will lead to alternative 
phenomenologies. Having made this choice for supersymmetry breaking, 
the location of each quark and lepton field in the fifth dimension is 
unique, up to a two-fold ambiguity,\footnote{
The two cases correspond to whether the five-plets, $F_i$, are in the 
bulk or on a brane.  We prefer the brane case, since only then does 
a unified prediction for $m_b/m_\tau$ follow, and quote predictions 
for this case in the conclusions.}
leading to definite predictions for both the superpartner spectrum and 
for lepton flavor violation.

The heart of our predictions rests on their being a single supersymmetry 
breaking boundary condition parameter, $\tilde{m}$, so that, even 
allowing for arbitrary $\mu$ and $B$ parameters in the Higgs potential, 
the entire superpartner spectrum depends on only $\tilde{m}$ and 
$\tan\beta$. Examples of this spectrum for two values of 
$(\tilde{m}, \tan\beta)$ are shown in Table~\ref{table:spectrum}. 
The range of $\tilde{m}$ and $\tan \beta$ is limited: $\tilde{m}$ cannot 
be much less than $200~{\rm GeV}$ from the experimental limit on the 
mass of the charged scalar tau, and should not be much more than about 
$500~{\rm GeV}$, since above this electroweak symmetry breaking rapidly 
becomes more fine tuned. The ratio of electroweak vacuum expectation 
values, $\tan\beta$, has a lower bound of about $3$ from the limit on 
the Higgs boson mass, and must be less than about $25$ to ensure that 
the mass squared for the right-handed scalar tau is positive. It is 
clear that the precision of the predictions for the entire spectrum 
in terms of these two parameters allows a very significant test of many 
aspects of the theory. In particular, the superpartners which reside 
on the brane, those in $T_3$ and $F_{1,2,3}$, have zero tree-level mass 
and get heavy only from renormalization group scaling. This is 
particularly clear in the sleptons, and is visible in the lightness of 
$\tilde{\tau}_{1,2}$ and $\tilde{l}_{1,2}$. These effects are also 
present in the squarks, although this is somewhat hidden by the gluino 
focusing effect. The mass eigenstates of the third generation squarks 
are not given by the soft mass-squared parameters for the helicity 
eigenstates because of the mixing induced by important $A$ parameters, 
which take the value $A_t = A_b = A_\tau = -\tilde{m}$ at the 
compactification scale, reflecting the Higgs residing in the bulk and 
the third generation on the brane. A detailed study of the superpartner 
and Higgs spectrum would thus not only measure $\tilde{m}$ and 
$\tan\beta$, but would also verify the location of each matter and 
Higgs field. 
\begin{table}
\begin{center}
\begin{tabular}{|c|c|c|}  \hline 
 $\tilde{m}$ & $300$ & $400$ \\
 $\tan\beta$ & $5$   & $10$  
\\ \hline
 $\tilde{g}$            & $699$ & $911$ \\
 $\tilde{\chi}^{\pm}_1$ & $251$ & $334$ \\
 $\tilde{\chi}^{\pm}_2$ & $427$ & $531$ \\
 $\tilde{\chi}^0_1$     & $130$ & $175$ \\
 $\tilde{\chi}^0_2$     & $251$ & $334$ \\
 $\tilde{\chi}^0_3$     & $417$ & $518$ \\
 $\tilde{\chi}^0_4$     & $422$ & $528$ 
\\ \hline
 $\tilde{q}$ & $701$ & $915$ \\
 $\tilde{u}$ & $675$ & $880$ \\
 $\tilde{d}$ & $602$ & $780$ \\
 $\tilde{l}$ & $209$ & $277$ \\
 $\tilde{e}$ & $317$ & $422$ 
\\ \hline
 $\tilde{t}_1$    & $425$ & $547$ \\
 $\tilde{t}_2$    & $619$ & $780$ \\
 $\tilde{b}_1$    & $563$ & $727$ \\
 $\tilde{b}_2$    & $601$ & $774$ \\
 $\tilde{\tau}_1$ & $106$ & $126$ \\
 $\tilde{\tau}_2$ & $214$ & $280$
\\ \hline
 $h$       & $118$ & $128$ \\
 $A$       & $552$ & $690$ \\
 $H^0$     & $553$ & $690$ \\
 $H^{\pm}$ & $558$ & $695$ 
\\ \hline
 $\alpha_s(M_Z)$ \{$\pm 0.003$\} & $0.119$ & $0.118$ \\
 $m_b(M_Z)$ \{$\pm 0.10$\}       & $3.37$  & $3.26$  
\\ \hline
 ${\rm Br}(\mu \rightarrow e\gamma)$                 & 
   $6\times10^{-12}$ & $8\times10^{-12}$ \\
 ${\rm Br}(\mu \rightarrow 3e)$                      & 
   $4\times10^{-14}$ & $5\times10^{-14}$ \\
 ${\rm Cr}(\mu \rightarrow e; {}^{48}_{22}{\rm Ti})$ & 
   $4\times10^{-14}$ & $5\times10^{-14}$ \\
 ${\rm Br}(\tau \rightarrow \mu\gamma)$              & 
   $1\times10^{-8}$  & $1\times10^{-8}$  
\\ \hline
\end{tabular}
\end{center}
\caption{Predictions for the superpartner spectrum, the Higgs spectrum, 
 gauge and Yukawa unification, and lepton flavor violating processes. 
 The predictions are for two representative values of $\tilde{m}$ and 
 $\tan\beta$, and all masses are given in GeV. Mass eigenvalues are 
 given for the gluino, $\tilde{g}$, the charginos, $\tilde{\chi}^\pm$, 
 the neutralinos, $\tilde{\chi}^0$, the squarks and sleptons of the 
 third generation, $\tilde{t}_{1,2}, \tilde{b}_{1,2}$ and 
 $\tilde{\tau}_{1,2}$, and the Higgs bosons, $h, A, H^0$ and $H^{\pm}$. 
 The mass of the lightest Higgs boson, $h$, includes one-loop radiative 
 corrections from top quarks and squarks.  For the first two generations 
 of squarks and sleptons the masses are shown for $\tilde{q}, \tilde{u}, 
 \tilde{d}, \tilde{l}$ and $\tilde{e}$ and do not include contributions 
 from electroweak $D$ terms.}
\label{table:spectrum}
\end{table}

The predictions of our theory for $\alpha_s(M_Z) $ from gauge coupling 
unification and the $b$ quark mass from Yukawa unification are
\begin{eqnarray}
 \alpha_s(M_Z) 
 &=& \left( 0.1327 - 0.0030 \, \ln\frac{\tilde{m}}{M_Z}
   -0.0019 \, \ln\frac{M_s}{M'_c} \right) \pm 0.003,
\\
 m_b(M_Z) 
 &=& \left( 3.62 - 0.022 \, \tan\beta 
   -0.026 \, \ln\frac{M_s}{M'_c} \right) \pm 0.1~{\rm GeV}.
\end{eqnarray}
In the above expression, the leading term is given first, the second 
term is the supersymmetric threshold correction and the third term is 
the unified scale correction; the uncertainty for $\alpha_s$ arises 
from unknown physics at and above $M_s$. These are remarkably precise 
predictions. The supersymmetric threshold corrections involve the two 
parameters $\tilde{m}$ and $\tan\beta$, have relatively small 
coefficients, and will be known once superpartner masses are measured. 
The assumption of strong 5D gauge interactions at $M_s$ leads to the 
prediction $M_s/M'_c \approx 200$. It is quite remarkable that the 
leading unified scale corrections are thus calculable and predicted, 
and move both predictions into agreement with experimental data. 
Using $\tilde{m}=400~{\rm GeV}$ and $\tan\beta = 10$, we obtain 
$\alpha_s(M_Z) = 0.118 \pm 0.003$ and $m_b(M_Z) = 3.3 \pm 0.1~{\rm GeV}$.
The prediction for the QCD coupling is in precise agreement with data, 
unlike the case of conventional supersymmetric grand unification, where 
large uncalculable unified threshold corrections are required.
We predict $m_b(M_Z) $ to be at the upper end of the presently 
allowed experimental region of $3.0 \pm 0.3~{\rm GeV}$.

Lepton flavor violation is an important experimental signal for 
conventional supersymmetric unified theories \cite{Barbieri:1994pv, 
Barbieri:1995tw}, providing soft supersymmetry breaking operators are 
local at the unified scale. The slepton non-universality arises from 
renormalization group scaling from the top quark Yukawa coupling above 
the unification scale. Uncertainties arise from two sources, flavor 
mixing matrices and the structure of the theory above the unified scale. 
Slepton non-degeneracy could also arise from scaling from the neutrino 
Yukawa coupling below the unified scale \cite{Borzumati:1986qx}. 
In this paper we have demonstrated that lepton flavor violation is 
also an important signature in our theory.  However, in contrast to 
4D theories, slepton non-degeneracy arises at tree level and is maximal. 
A heavy top quark results from $T_3$ being located on the brane, so that, 
at tree level, $\tilde{\tau}_R$ does not feel the supersymmetry breaking 
boundary condition and is massless. By contrast $T_1$ must be located 
in the bulk, otherwise gauge boson exchange leads to too large a proton 
decay rate, so that $m_{\tilde{e}_R} = \tilde{m}$ at the compactification 
scale. This maximal slepton non-degeneracy leads to larger rates for 
lepton flavor violation in our 5D theory than in conventional 
supersymmetric unified theories, and furthermore reduces the 
uncertainties of the signal.  While there is still a dependence on 
the flavor mixing matrices, the uncertainties associated with the 
generation of the slepton non-degeneracy is removed. The first immediate 
consequence is that $T_2$ must be located with $T_1$ in the bulk, 
to avoid a large non-degeneracy between $\tilde{e}$ and $\tilde{\mu}$, 
and the three $F_i$ must have a common location to avoid too large 
lepton flavor violation from diagrams involving non-degeneracies 
in both left and right-handed sleptons.  With $F_i$ on the brane, 
the branching ratios for flavor violating lepton decays are found 
to be close to the present experimental limits 
\begin{eqnarray}
 && {\rm Br}(\mu \rightarrow e\gamma) 
 \simeq 3 \times 10^{-11} 
     \left(\frac{200~{\rm GeV}}{\tilde{m}}\right)^4
     \left(\frac{|W^e_{\tau\mu}|}{0.04}\right)^2 
     \left(\frac{|W^e_{\tau e}|}{0.01}\right)^2 
     \left(\frac{\tan\beta}{5.0}\right)^2,
\\
 && {\rm Br}(\mu \rightarrow 3e) 
 \simeq {\rm Cr}(\mu \rightarrow e; {}^{48}_{22}{\rm Ti})
 \simeq 2 \times 10^{-13} 
     \left(\frac{200~{\rm GeV}}{\tilde{m}}\right)^4
     \left(\frac{|W^e_{\tau\mu}|}{0.04}\right)^2 
     \left(\frac{|W^e_{\tau e}|}{0.01}\right)^2 
     \left(\frac{\tan\beta}{5.0}\right)^2,
\\
 && {\rm Br}(\tau \rightarrow \mu\gamma) 
 \simeq 5 \times 10^{-8} 
     \left(\frac{200~{\rm GeV}}{\tilde{m}}\right)^4
     \left(\frac{|W^e_{\tau\mu}|}{0.04}\right)^2 
     \left(\frac{|W^e_{\tau\tau}|}{1.0}\right)^2 
     \left(\frac{\tan\beta}{5.0}\right)^2.
\end{eqnarray}
Once $\tilde{m}$ and $\tan\beta$ are determined from the superpartner
spectrum, observation of these decay modes would measure the two
independent flavor mixing angles of the lepton mixing matrix $W^e$. 
Indirect evidence for our theory would follow if the size of this 
intergenerational mixing is comparable to that measured in the quark 
sector. To go further would require a more detailed theory of flavor 
than we have given here. 

We stress that large lepton flavor violation is a generic signature 
in any KK grand unified theory where the supersymmetry breaking 
reflects the structure of matter location in extra dimensions. 
The matter locality breaks $U(3)$ flavor symmetry, leading to squark 
and slepton mass matrices of the general form of Eq.~(\ref{eq:sq-sl-m2}) 
dictated by the flavor symmetry of the 5D gauge interactions. The low 
energy superparticle spectrum then reveals characteristic features 
reflecting the structure of this new flavor symmetry, irrespective 
of how supersymmetry is broken. Therefore, lepton flavor violation 
probes all supersymmetry breaking schemes which gives soft operators 
local up to the compactification scale, and the detailed superpartner 
spectroscopy will uncover the geometry of matter fields in extra 
dimensions and help discriminate between various possibilities for 
supersymmetry breaking.

Proton decay can occur in our theory via the bulk gauge interactions 
of the $X$ gauge boson through flavor mixing matrices. However, the 
resulting partial lifetime $\tau(p \rightarrow K^+ \bar{\nu}) \approx 
10^{37 \pm 2}~{\rm years}$ is probably too long to be reached by 
future experiments. Proton decay can also be mediated by brane 
kinetic operators, and these we estimate to give a lifetime of 
about $10^{34}~{\rm years}$.  The structure of the final states is 
very rich with comparable branching ratios to $e^+\pi^0, \mu^+\pi^0, 
e^+K^0, \mu^+K^0, \pi^+\bar{\nu}$ and $K^+\bar{\nu}$.  Although the 
uncertainty in the lifetime is large as the coefficient for the 
relevant brane operator is not predicted, the branching ratios are all 
given in terms of a single unknown mixing parameter.  Since gauge boson 
mediated proton decay does not involve an exchange of superparticles, 
these results are completely independent of supersymmetry breaking.

As stressed above, the requirements of a large top quark mass and 
proton longevity require a separation in location between the top 
quark and the up quark --- $T_3$ must be on the brane and $T_1$ in the 
bulk. Introducing supersymmetry breaking by a twist in the boundary 
condition, this requires that $F_{1,2,3}$ share a common location, 
to avoid too large $\mu \rightarrow e$ and $\tau \rightarrow \mu$ 
transition rates. This immediately leads to a prediction of 
large neutrino mixing angles following from the see-saw generated 
neutrino mass matrix; both atmospheric and solar neutrino oscillations 
should result from large mixing angles. The Majorana mass for the 
right-handed neutrino arises from a brane-localized operator, and 
must have a size which is suppressed relative to the cutoff scale 
of our effective 5D theory. This suggests that the right-handed 
neutrino mass is protected by some symmetry, which we take to be 
the $U(1)_X$ remnant of $SO(10)$. The breaking of $U(1)_X$ gauge 
symmetry leads not only to right-handed neutrino masses, but also, 
when supersymmetry is broken, to the $\mu$ and $B$ parameters of 
the Higgs potential.

We have seen that a $U(1)_R$ symmetry is a critical feature of our
theory, yet this symmetry is clearly broken by the supersymmetry
breaking operators generated by the boundary conditions. Ultimately, 
the supersymmetry breaking will be spontaneous, arising from the 
vacuum expectation value of a field in the 5D supergravity multiplet. 
It therefore seems natural to assume that all breaking of $U(1)_R$ is 
also spontaneous, in which case there is an $R$ axion. This $R$ axion 
has a QCD anomaly and may therefore solve the strong $CP$ problem.  
We also find that its decay constant may be in an interesting range 
for the axion to be dark matter.

We have found that $SU(5)$ unification in 5D offers many advantages 
over unification in 4D. While we have not addressed the origin of 
radius stabilization or matter localization, our effective field 
theory is remarkably simple; for example, the only non-trivial $SU(5)$ 
multiplets beyond the gauge multiplet are five-plets and ten-plets. 
Our theory is sufficiently constrained that it offers several avenues 
for experimental tests.  We expect the first direct experimental 
signal for our theory to be the observation of events containing two 
``stable'' charged particles at the Tevatron or at LHC. These scalar 
taus have opposite charges if they arise from Drell-Yan production, 
but have equal probability of like charge and opposite charge 
combinations if they are produced from squark and gluino decays. 
It is likely that these charged scalar taus decay cosmologically to 
neutral axinos, which may contribute to dark matter of the universe.

\section*{Acknowledgments}

Y.N. thanks the Miller Institute for Basic Research in Science 
for financial support.  This work was supported in part by the Director, 
Office of Science, Office of High Energy and Nuclear Physics, of the U.S. 
Department of Energy under Contract DE-AC03-76SF00098, and in part 
by the National Science Foundation under grant PHY-00-98840.

\end{document}